\documentclass[11pt]{article}
\usepackage{amsmath}
\usepackage{amssymb}
\usepackage{graphics}
\usepackage{hyperref}
\hypersetup{colorlinks=true, linkcolor=blue, citecolor=blue}

\newcommand{\tr}{\operatorname{tr}}

\begin{document}

\title{Non-Perturbative Groundstate of High Temperature Yang-Mills Theory}
\author{d yamada\footnote{{\tt dyamada@protonmail.ch}}
        \bigskip
        \\
       {\it Shizukari Danchi Center},
        \smallskip
       \\
         {\it Shizukari 5-3, Oshamanbe-Cho, Hokkaido, 049-5141 Japan}
%          \bigskip
}
\date{}
%no argument gives no date. Getting rid of this line above puts today's date.
\maketitle

%%%%%%%%%%%%%%%%%%%%%%%%%%%%%%%%%%%%%%%%%%%%%%%%%%%%%%%%%%%%%%%%%%%%%
\begin{abstract}
%%%%%%%%%%%%%%%%%%%%%%%%%%%%%%%%%%%%%%%%%%%%%%%%%%%%%%%%%%%%%%%%%%%%%
  The effective potential of Yang-Mills theory at high temperature
  derived by Gross, Pisarski, Yaffe and Weiss is critically reexamined
  and it is argued that the groundstate of the potential at  $\langle A_0 \rangle = 0$ is invalid,
  due to the infrared divergence of the Matsubara zero mode.
  This suggests that the thermal groundstate is dominated by infrared non-perturbative effects.
  Lattice simulations are carried out and the field $A_0$ in the static gauge
  is observed to acquire nonzero, non-perturbative expectation values at high temperatures.
  A consequence is that thermal perturbation theory is inconsistent with the non-perturbative groundstate
  and it cannot account for all of the contributions to a thermodynamic quantity at any temperature.
  Related issues, including dimensional reduction and confinement, are also discussed.
\end{abstract}

\bigskip

%%%%%%%%%%%%%%%%%%%%%%%%%%%%%%%%%%%%%%%%%%%%%%%%%%%%%%%%%%%%%%%%%%%%%
\tableofcontents
%%%%%%%%%%%%%%%%%%%%%%%%%%%%%%%%%%%%%%%%%%%%%%%%%%%%%%%%%%%%%%%%%%%%%

%\pagebreak

%%%%%%%%%%%%%%%%%%%%%%%%%%%%%%%%%%%%%%%%%%%%%%%%%%%%%%%%%%%%%%%%%%%%%
\section{Introduction}\label{sec:intro}
%%%%%%%%%%%%%%%%%%%%%%%%%%%%%%%%%%%%%%%%%%%%%%%%%%%%%%%%%%%%%%%%%%%%%
That QCD is the theory of strong interaction comes most unequivocally from
the zero temperature high energy regime.
Thanks to asymptotic freedom, the theory is under perturbative
control in the regime and important contributions to observables come from small
fluctuations around the trivial groundstate, $\langle A_\mu\rangle = 0$.
The perturbation theory is eminently successful in describing the data from various high energy experiments.

Thus it is quite natural to expect that the theory at high temperature -- much higher than the QCD scale --
is amenable to perturbation theory expanded around the same trivial thermal groundstate/equilibrium,
$\langle A_\mu\rangle = 0$.
In particular, the ``scalar field'' of the theory, $A_0$, is supposed to have
the trivial groundstate expectation value --
this is despite the fact that the spacetime symmetry no longer demands so at finite temperature.
This supposition appears to borne out in the one-loop effective potential of Yang-Mills theory
at high temperature; the classic result obtained by Gross, Pisarski and Yaffe \cite{Gross:1980br}, and by Weiss~\cite{Weiss:1980rj}.
The GPYW potential is a function of the diagonal $A_0$ components and it 
confirms that the minimum is located at the origin; that is, $\langle A_0\rangle = 0$.

This result, however, is more ambiguous than it seems.
The operator $A_0$ is gauge covariant, so it automatically averages to zero without gauge fixing.
When the gauge is fixed, we should expect the average be generally gauge dependent
unless some kind of symmetry forces it otherwise.
This implies that different gauges may yield different expectation values and for some,
$\langle A_0\rangle$ could happen to vanish.
In Section~\ref{sec:effS}, I will show that implicit supersymmetry actually requires the gauge
independence of $\langle A_0\rangle$ but only at one-loop level.
Anishetty, in Reference~\cite{Anishetty:1981tm}, claimed that the two-loop potential
developed a nontrivial minimum away from the origin; i.e., $\langle A_0\rangle \neq 0$.
Subsequently it was found that the two-loop potential and the location of the minimum were
gauge dependent, setting the debate over whether $A_0$ develops physically relevant vacuum expectation value.
This question of the ``$A_0$ condensate,'' in my opinion,  has not reached a clear consensus;
see References \cite{Belyaev:1989yt,Enqvist:1990ae,Borisenko:1994jn,Borisenko:2020dej},
especially the review article~\cite{Borisenko:1994jn}.
Gauge dependent or not, $\langle A_0\rangle \neq 0$ is highly consequential
because it puts whole thermal perturbation theory on a shaky ground;
the theory {\it assumes} $\langle A_0\rangle = 0$ as its groundstate, so a correction to
this assumption would interfere with the established perturbative results.

There is also vagueness in the derivations of the effective potential.
The computations are always carried out by adopting a constant, diagonal background field for $A_0$.
In the literature, it is vague in what justifies this particular form of background.
This clearly is not a choice of gauge because a background field does not transform under
{\it true} gauge transformations.
In Section~\ref{sec:critique}, I will clarify that it rather is a fixing choice for the {\it formal} background
gauge symmetry enjoyed by the system under the background field gauge.
Then I will explain how this background field gauge is not adequate in extracting physical information from the effective potential,
due to the arbitrary gauge dependence mentioned above.
Instead, the importance of the {\it static gauge} is emphasized throughout this paper.

Nadkarni, in Reference~\cite{Nadkarni:1982kb}, also pointed out the problem with the vanishing $A_0$ expectation value.
He found that the electrostatic component of the two-loop gluon self-energy,
$\Pi_{00}$, suffered from gauge dependence and infrared divergences in the limit of zero external momentum.
He did not emphasize but this means that the second derivative of the effective potential at the origin is gauge
dependent and divergent at two-loop level.
This led Nadkarni to speculate on the possibility of $\langle A_0\rangle \neq 0$ \cite{Nadkarni:1986as}.

In Section~\ref{sec:critique} of this work, I will present compelling arguments
that the one-loop GPYW potential is infrared divergent at the origin due to invalid infrared regularization.
This manifests itself as the discontinuity in the third derivative of the potential at the origin.
The discontinuity was noticed long ago \cite{Enqvist:1990ae}, but I will go further and
argue that the potential yields no physical information as its groundstate is ill-defined.
Put it differently,
the effective potential is invalidated by the infrared divergence even at one-loop level
and the conclusion $\langle A_0\rangle = 0$ cannot be drawn;
{\it the true thermal equilibrium is likely dominated by the infrared non-perturbative effects}
suggested by the divergences.
I will confirm this on the lattice in Section~\ref{sec:lattice} by directly observing $\langle A_0\rangle \neq 0$
in the static gauge.

This is not a radical claim but is the quite conspicuous elephant in the room.
It has long been known that thermal perturbation of non-abelian Yang-Mills theory is pathological.
Linde argued in Reference~\cite{Linde:1980ts} that {\it at best}, the perturbation makes sense only up to
certain orders, depending on the quantities of interest.
For instance, the free energy is up to the order of $g^5$, $g$ the Yang-Mills coupling.
At higher orders, the expansion breaks down as uncontrollable number of diagrams contribute to the same orders;
a phenomenon widely attributed to the non-perturbative infrared divergences of the theory.

Another perspective on the pathology was pointed out by GPY themselves~\cite{Gross:1980br}.
They note that the theory at high enough temperature can be
described by an effective three dimensional theory.
This comes about as this \cite{Appelquist:1981vg}: In the imaginary time formalism of the theory, the heavy nonzero
Matsubara modes decouple \cite{Appelquist:1974tg}
and can be integrated out; the remaining degrees of freedom are
the fields' zero modes and by definition, they are time-independent and the theory effectively lives
on three spatial dimensions.
The point is, this 3D theory is highly non-perturbative and confining
-- confining from the standpoint of three dimensions --
so the perturbation theory is destined to break down.

The hope has been that the perturbation series provides most important information before it loses control
and the incalculable leftover may be safely ignored.
This is the best case scenario of Linde's.
His worst, in a sense, is that
high temperature perturbation theory does not make sense due to an unstable perturbative groundstate.
In this case, all the perturbative computations that have been pressed forward would be on a wrong groundstate.
This grave possibility has not been ruled out convincingly.
In fact, it takes asymptotically large temperatures to make the perturbation series of
the free energy viable~\cite{Arnold:1994ps}.
In addition, some perturbatively computed quantities are known to agree with the lattice data
only at unrealistically large temperatures.
See, for example, Reference~\cite{Kajantie:1997tt}.

This paper is to point out the breakdown of thermal perturbation from the get-go
at one-loop level in the effective potential, by focusing on the expectation value of $A_0$.
Lattice simulations are carried out to demonstrate the non-perturbative nature of the thermal
groundstate through the non-vanishing expectation value.
The table of contents outlines the structure of discussion.

%%%%%%%%%%%%%%%%%%%%%%%%%%%%%%%%%%%%%%%%%%%%%%%%%%%%%%%%%%%%%%%%%%%%%
\section{Three Dimensional Effective Theory}\label{sec:effS}
%%%%%%%%%%%%%%%%%%%%%%%%%%%%%%%%%%%%%%%%%%%%%%%%%%%%%%%%%%%%%%%%%%%%%
I start by clarifying some notions in constructing a 3D effective theory of high temperature Yang-Mills theory.
The main purpose of this section is to introduce the static gauge and emphasize the importance of it.
I will also explain how a subtlety impedes straightforward construction of a Wilsonian effective action,
leaving the matching method the only viable option.
As a result, I stress that the 3D effective theory is not a derived entity but a guess.

Later in this section, I will re-derive the GPYW potential in the process of computing the one-loop, non-derivative
scalar potential of the 3D effective theory.
I will show that the one-loop result is gauge independent.

Throughout this work, I use Euclidean signature and the convention $\tr\big(t^at^b)=(1/2)\delta^{ab}$
where $t^a$ are the Hermitian generators of the gauge group in the defining representation.

%%%%%%%%%%%%%%%%%%%%%%%%%%%%%%%%%%%%%%%%%%%%%%%%%%%%%%%%%%%%%%%%%%%%%
\subsection{Static Gauge and Dimensional Reduction}

Given a momentum scale, integrate over the degrees of freedom with momenta above the scale and construct a theory
that describes the lower scale physics -- this is the idea of the Wilsonian effective action.
In the lower scale theory, the dynamics above the scale is suppressed and its effects
are encoded in the coefficients of the operators in the effective Lagrangian.
The effective Lagrangian generally contains infinitely many terms but only a handful of
terms should be ``relevant'' in a useful effective theory.

In the context of Yang-Mills theory at high temperature, one evident scale is the temperature, $T$.
In addition, the imaginary time formalism offers the separation of the fields into
zero and nonzero Matsubara modes.
From the perspective of three spatial dimensions $T$ separates the light zero modes and heavy nonzero modes,
motivating us to integrate over the heavy modes and construct a three dimensional effective theory
of zero modes.

To most of the readers, it is obvious that after integrating out the heavy modes,
the 3D effective Lagrangian assumes the form
\begin{equation}\label{eqn:L3D}
  \mathcal{L}_\text{3D} = \frac{1}{4}\big(F_{ij}^a\big)^2
  + \frac{1}{2}\big(D_iA_0^a\big)^2 + V_\text{3D}(A_0) + \cdots
  \;,
\end{equation}
where the indices $i,j$ run through three spatial dimensions while the index $a$ through the color space, $A_0$ is the adjoint scalar field
descended from the zeroth component of the 4D gauge field, $D_i$ is the 3D covariant derivative,
$V_\text{3D}$ is the non-derivative potential of the scalar field and the dots represent other terms
including higher derivative operators.
This electrostatic theory, EQCD$_3$, has a highly constrained form and the reader knows that gauge invariance does not allow otherwise.
But what gauge invariance? In order to compute any quantum effects, one must fix the gauge, i.e., the gauge symmetry
must be explicitly broken to integrate out the heavy modes.
Then there is no guarantee that the resulting effective Lagrangian takes such a special form.
Moreover, the gauge symmetry has been significantly reduced from the 4D to 3D symmetry.
How did it happen? The dimensional reduction is not as intuitive as it seems.

A particular gauge can help save the intuition.
\\
\\%
{\bf Problem with Defining the Zero Mode}

Let me assume the imaginary time formalism and split the original four dimensional gauge field, $A_\mu(t,\vec x)$, into
zero and nonzero Fourier modes:
\begin{equation}\label{eqn:split}
  A_\mu = \hat{A}_\mu + A_\mu'
  \quad
  \text{where}\quad \partial_0 \hat{A}_\mu = 0 \quad\text{and}
  \quad
  \int_0^{1/T} \! dt \, A_\mu' = 0
  \;.
\end{equation}
The aim is to integrate over the field $A_\mu'$.
This split, however, is ill-defined without gauge fixing
because it is spoiled by gauge transformations as a spacetime-dependent
transformation parameter $\epsilon(t,\vec{x})$ mixes up the modes:
\begin{equation} \label{eqn:gaugeTransf}
  \delta A_\mu^a = \frac{1}{g}\partial_\mu\epsilon^a(t,\vec{x})
    + f^{abc}\epsilon^b(t,\vec{x})\big\{ \hat{A}_\mu^c(\vec x) + {A_\mu'}^c(t,\vec x)\big\} \;,
\end{equation}
where $f^{abc}$ are the group structure constants with color indices.
This calls for gauge fixing and implies that
dimensional reduction by integration is an inherently gauge dependent procedure.
\\
\\%
{\bf The Static Gauge}

I want a convenient gauge that allows for the clean split; moreover, I want it preserve the 3D spatial gauge invariance that
imposes the stringent organizing principle.
The latter, however, is impossible because the gauge must be fixed completely for general quantization.
Now, a gauge fixed Lagrangian has a global symmetry -- the BRST symmetry -- and this is what should be looked at.
This discussion involving the BRST symmetry is straightforward but longer,
so relegated to Appendix~\ref{app:BRST}.

Here we take a slightly heuristic approach by considering partial gauge fixing and this is 
the {\it static gauge} \cite{Weiss:1980rj,Nadkarni:1982kb,DHoker:1981bjo}
\begin{equation}\label{eqn:staticgauge}
  \partial_0 A_0 = 0
  \quad \Leftrightarrow \quad
  A_0' = 0
  \;.
\end{equation}
I call this the static gauge and a static gauge is referred to as the static gauge with the remaining gauge
symmetry fixed by a gauge.
Here is a list of some nice features about this gauge:
The Lorentz (Euclidean) symmetry in the time direction is already broken
by the introduction of temperature; thus, this gauge breaks no further spacetime symmetry.
Also as indicated in Equation~(\ref{eqn:staticgauge}), this is the temporal gauge imposed on the $A_\mu'$ field. 
As such, the ghosts associated with this gauge become irrelevant in integrating out
the heavy fields \cite{Nadkarni:1982kb,Williams:2002dw}.
Furthermore, the expectation value, $\langle A_0 \rangle$, in the static gauge is an
order parameter of the $\mathbb{Z}_N$ symmetry, giving the static gauge a physical significance
in terms of the $A_0$ expectation value.
I will further explain this last point later.

And the static gauge leaves the time-independent gauge freedom, under which the split fields transform as%
\footnote{One might find the big parentheses in the right-hand side of Equation~(\ref{eqn:splittransf}) rather arbitrary
  by associating the derivative term to the time-independent sector. But the derivative term is the pure gauge part
  and the pure gauge should certainly be in the time-independent sector.}
\begin{equation}\label{eqn:splittransf}
  \delta\big( \hat{A}_\mu + A_\mu' \big) = \bigg( \frac{1}{g}\partial_\mu\epsilon^a(\vec{x})
  + f^{abc}\epsilon^b(\vec{x})\hat{A}_\mu^c(\vec x) \bigg)
  + f^{abc}\epsilon^b(\vec{x}){A_\mu'}^c(t, \vec x) \;.
  \;
\end{equation}
Thus the action is invariant under a formal transformation law%
\footnote{One should note that the gauge transformation acts as in Equation~(\ref{eqn:splittransf}) and not
  as in Equation~(\ref{eqn:3Dtransf}). The latter is a formal transformation that will become 3D gauge transformation
after integrating out the heavy modes.}
\begin{eqnarray}\label{eqn:3Dtransf}
  \delta \hat{A}_0^a(\vec x) &=& f^{abc}\epsilon^b(\vec{x}){\hat{A}_0}^c(\vec x)
  \nonumber\\
  \delta \hat{A}_i^a(\vec x) &=& \frac{1}{g}\partial_i\epsilon^a(\vec{x}) + f^{abc}\epsilon^b(\vec{x})\hat{A}_i^c(\vec x)
  \nonumber\\
  \delta {A_i'}^a(t,\vec x) &=& f^{abc}\epsilon^b(\vec{x}){A_i'}^c(t,\vec x)
  \;.
\end{eqnarray}
We see that the split is well-defined under the time-independent gauge transformation.
Note that the $\hat A_0$ and $A_i'$ fields are transforming as adjoint matters while
$\hat A_i$ as a gauge field.

Without fixing the gauge further, the integration over the $A_i'$ field can be carried out as its propagator is defined.
The other fields are treated as either background or external fields \cite{Abbott:1981ke}.
When treated as background, the $A_i'$ propagator and the vertices include factors of the background fields
and the complexity generally prevents computations except for very simple few cases.
Thus, in practice, one usually treats $\hat A_\mu$ as external fields and expand with respect to them.
(They appear as external ``legs'' of diagrams.)
The integration measure of $A_i'$ should remain invariant under a simple linear transformation
in Equation~(\ref{eqn:3Dtransf}), so that the result of the integration over $A_i'$
is invariant under the 3D gauge transformation: the first two lines of Equation~(\ref{eqn:3Dtransf}).
The resulting 3D effective Lagrangian, therefore, takes the form of Equation~(\ref{eqn:L3D}),
after dropping the hats on the fields.

Notice the unique role played by the static gauge in dimensional reduction:
it breaks the 4D gauge symmetry to 3D but preserves the latter while allowing for the well-defined
heavy modes and their integration.
The importance of the static gauge will be further discussed in the context of determining $\langle A_0 \rangle$.
\\
\\%
{\bf The Fatal Subtlety}

The 3D effective theory discussed so far is still incomplete.
At a high but finite temperature, there is a subtlety that turns out to be fatal.

The subtlety is the hard spatial momenta of the zero modes.
They must also be integrated out down to somewhat below $T$ for a consistent Wilsonian low energy effective action.%
\footnote{Hard zero modes can excite heavy modes making the theory essentially four dimensional
  and they make all infinitely many terms in the effective action important.}
Now the problem is that this integration over the hard zero modes cannot be done in
a 3D gauge, BRST or Euclidean invariant fashion.
Integration over the zero modes requires the 3D gauge fixing that breaks the gauge symmetry.
No known choice of such a gauge would lead to a BRST transformation that preserves the split of the fields
into two regions of momentum.
And such split violates the 3D Euclidean invariance.
Therefore, integration over the hard zero modes
results in a Wilsonian effective theory that is not a gauge theory, like chiral perturbation theory,
but unlike $\chi$PT there is no symmetry principle to restrict the form of the action;
this would complicate the theory in an impractical manner.

Conclusion: A viable 3D Wilsonian effective action of high temperature Yang-Mills theory
cannot be constructed by actually integrating out the hard modes,
except in the limit of infinite temperature.
\\
\\
\\
\\
{\bf The Matching Method}

This conclusion leaves us only with the matching method \cite{Braaten:1995jr,Manohar:2018aog};
it is not just a matter of convenience but is an essential procedure to construct a high temperature effective theory.

First in this method, relevant degrees of freedom in low energy should be decided (or guessed) and it is useful to have
an organizing principle, such as symmetries.
In our case, guided by the static gauge and its dimensional reduction,
a 3D gauge theory with an adjoint scalar field is a reasonable candidate to try.
So write terms in this theory consistent with the gauge symmetry with unknown coefficients.
Here, the terms are not restricted to renormalizable ones.
Contributions to an amplitude from non-renormalizable terms are suppressed by the powers of a factor
$p/T$, where $p$ is a typical low energy momentum scale.
Thus the number of terms included in the action depends on the required accuracy of the computation.
%\footnote{The scale of the 3D effective theory is $p=g^2T$, the 3D coupling, so the expansion is
%  in terms of $g^2$.
%  This looks like a perturbation series, but it is not; it is a momentum expansion series.}
Then low energy observables are computed in the effective theory and they are matched to the counterparts computed in the full
theory, or they can also be matched to experiment or lattice data.
The matching determines the coefficients in the effective theory.

The gauge fixing and ultraviolet regularization may be implemented in the full and effective theories independently
because the differences are adjusted in the choice of the coefficients.%
\footnote{Dimensional regularization is a popular choice, but this regularization must be used both in the full and effective theories.
  The reason is that it regulates both UV and IR divergences. For instance, some scaleless integrals are regulated to vanish
  due to the cancellation between the UV and IR divergences \cite{Manohar:2018aog}. 
  This is a completely wrong thing to do, because $\epsilon$
  in $d-2\epsilon$ is required to be positive to regulate UV divergences and negative for IR divergences; so the $\epsilon$
  is positive and negative at the same time. This is acceptable as long as one commits exactly the same wrong doing
  in both theories.
  The IR sectors are identical so the effects of wrong IR regularizations cancel out in the matching.}
In the effective theory, the loop momenta run up to infinity, but the inadequate contributions from hard momenta
are adjusted in the matching coefficients to produce the same low energy results as the full theory or other data,
to the required accuracy \cite{Manohar:2018aog}. 
There is no need to worry about exciting the heavy modes because there are none.
We have {\it decided} to try the 3D gauge theory as an effective theory.
It has no direct connection to the full theory, but we are {\it making} the connection by the matching.
In fact, the difference in the ultraviolet structures leads to the differences in the field renormalizations
and anomalous dimensions.
This implies that the fields in the full and effective theories are separate entities; one is not descending from the other.
This is a setback in the matching method: the evolution of the degrees of freedom from high to low
scales is in the black box (think again $\chi$PT).
However, given a required accuracy, only a finite number of observables are needed for the matching and the rest
are subject to the prediction; this can be a viable theory.

I want to stress that the 3D effective theory as a low energy description of high temperature Yang-Mills theory
is not the conclusion directly derived from the full theory but just an educated guess,
originally based on the decoupling argument \cite{Appelquist:1981vg}.%
\footnote{In general, it requires special renormalization schemes for the decoupling to occur \cite{Manohar:2018aog}.
  In high temperature Yang-Mills theory, the heavy modes do not decouple even with a carefully chosen
  renormalization scheme in infinite temperature limit \cite{Landsman:1989be}.}
Whether it is a useful effective theory or not must be judged in practice -- and it has not been very useful
in the context of QCD, mainly because it is expected to be non-perturbative and difficult to solve anyway.
It is completely possible that entirely different degrees of freedom are required for a useful 3D description.
After all, it is well-known that the 3D theory confines; then we should expect radical evolution of
degrees of freedom at low energy.

%%%%%%%%%%%%%%%%%%%%%%%%%%%%%%%%%%%%%%%%%%%%%%%%%%%%%%%%%%%%%%%%%%%%%
\subsection{One-Loop 3D Potential}
I am going to derive the non-derivative 3D potential of the adjoint $\hat A_0$ scalar,
$V_\text{3D}$ in Equation~(\ref{eqn:L3D}), by integrating out the heavy modes at one-loop level.
This should correspond to the potential at an arbitrarily large temperature.
In the process the GPYW potential is derived as a by product.

The 4D Lagrangian is
\begin{equation}\label{eqn:4DSUSYL}
  \mathcal{L} = \frac{1}{4}\big(F_{\mu\nu}^a\big)^2 + \mathcal{L}_\text{GF} + \mathcal{L}_\text{ghost}
  + i\sum_{f=1}^{N_f} \bar{\psi}_f^a \gamma_\mu D_\mu^{ab} \psi_f^b
  \;,
\end{equation}
where gauge fixing and corresponding ghost terms are included; see Nadkarni \cite{Nadkarni:1982kb}
or Appendix~\ref{app:BRST} for
the explicit expressions of these terms in static gauges.
I have also included $N_f$ adjoint Dirac fermions with $D_\mu^{ab}:=\delta^{ab}\partial_\mu - gf^{acb}A_\mu^c$,
for the reason that will become clear shortly.

Split the gauge field of the 4D Lagrangian according to Equation~(\ref{eqn:split}).
I am going to treat $\hat{A}_\mu$ as a background field; the narrow focus on a non-derivative potential makes
this treatment computationally doable.
For the background $\hat{A}_0$, I choose a constant, diagonal form.
As mentioned in the Introduction, this very special form needs justification.
I can set the field to $c$-number constant, since the derivative terms are not included in the potential.
Now, I know that the potential {\it will} organize itself in
traces of $\hat{A}_0$, because I have devised the Lagrangian~(\ref{eqn:L3D})
in a manifestly gauge invariant form using the static gauge.
Then the relevant background fields for this computation are the $\hat{A}_0$ eigenvalues,
allowing me to set $\hat{A}_0$ diagonal -- this is not a choice of gauge
but a choice afforded by the static gauge.

The plan is to use such $\hat{A}_0$ as a background and obtain the functional determinant by integrating out
the quadratic heavy modes; then translate the eigenvalues back into traces
of $\hat{A}_0$ and promote them to the $q$-number operators in the effective Lagrangian.
\\

Now, for a moment, let us imagine something entirely different.
Consider the case where the compactified dimension is one of the spatial directions, instead of the time direction,
and choose periodic boundary condition for the fermions.
Also let us suppose that we are integrating out all the modes, including the zero modes.
Then the theory is (one-shell) supersymmetric, provided that $N_f=1/2$, meaning
the fermion is either a Majorana or Wyle.
We know, in such a theory, that the vacuum-to-vacuum diagrams exactly cancel among themselves.
At one-loop level, in particular, the purely bosonic contributions are exactly canceled by the purely fermionic one.
This implies that the bosonic contributions, including the ghosts,
can be obtained by computing a simple fermionic loop with an appropriate minus sign and $N_f=1/2$.
This incidentally proves that the one-loop bosonic result is gauge independent, because the fermionic counterpart has nothing
to do with the gauge fixing.
(At higher loops, all the bubble diagrams involve gluon propagators, so they are generally gauge dependent,
before canceling among themselves.)
One can freely reinterpret the bosonic result in the context of finite temperature;
then subtraction of the zero mode contributions yields what we need.
\\

I package the $SU(N)$ gauge fields into $N\times N$ traceless matrices. Because my focus now is
only on the $\hat{A}_0$ potential, I turn off the fields $\hat{A}_i$.
Then the split field is
\begin{equation}
  A_\mu(x) = \hat{A}_0 \delta_{0\mu} + A_\mu'(x)
  \quad\text{where}\quad
  (\hat{A}_0)_{mn} = \lambda_{m}\delta_{mn}
  \quad\text{with}\quad
  \sum_{n=1}^N\lambda_n=0
  \;.
\end{equation}
In the original 4D Lagrangian, split the field as above and collect the terms quadratic in fermions to get
\begin{equation}
   2{\sum_{f,m,n}}' \big( \bar{\psi}_f \big)_{mn}^*
  \big( i\gamma_\mu\partial_\mu - g\lambda_{mn}\gamma_0 \big)
  \big( \psi_f \big)_{mn}
  \;.
\end{equation}
where the prime on the sum indicates the omission of the case $m=n=N$ to account for
the traceless nature of the fields, I have defined $\lambda_{mn} := \lambda_m - \lambda_n$ and
the factor of $2$ in front is related to my convention $\tr\big(t^at^b)=(1/2)\delta^{ab}$.
Carrying out the functional integral in the momentum space, we have
\begin{eqnarray}\label{eqn:fermiDet}
  &&{\prod_{f,m,n}}'\det\big( \gamma_\mu p_\mu - g\lambda_{mn}\gamma_0 \big)
  = {\prod_{f,m,n}}'{\det}^{1/2}\big( -p_\mu p_\mu + 2g\lambda_{mn}p_0-g^2\lambda_{mn}^2 \big)\mathbf{1}_{4\times 4}
  \nonumber\\
  && = {\prod_{m,n}}'{\det}^{2N_f}\big[ (p_0-g\lambda_{mn})^2+p_ip_i \big]
  \;,
\end{eqnarray}
where I used the usual ``$\gamma_5$-trick'' to square and diagonalize the operator
in the four dimensional representation space of the Clifford algebra.
According to the SUSY argument of the previous paragraph, the bosonic contribution is
\begin{equation}\label{eqn:bosonDet}
  {\prod_{m,n}}'{\det}^{-1}\big[ (p_0-g\lambda_{mn})^2+(p_i)^2 \big]
  \;.
\end{equation}

What appears in the 3D effective Lagrangian~(\ref{eqn:L3D}) is minus the logarithm of Equation~(\ref{eqn:bosonDet})
where the minus sign accounts for the one in the path integral weight factor $e^{-S}$.
I also need to divide the expression by the spatial volume $L_s^3$, since the action is the spatial
integral over the Lagrangian density.
So we have the constant $c$-number potential
\begin{eqnarray}\label{eqn:cV}
  V_\text{3D} &=& \frac{1}{L_s^3}{\sum_{m,n}}'\tr\ln\big[ (p_0-g\lambda_{mn})^2+(p_i)^2 \big]
  \nonumber\\
  &=& \frac{2T^3}{\pi^2}{\sum_{m,n}}'\bigg[\bigg\{ -\frac{\pi^4}{90} + \frac{\pi^2 [g\lambda_{mn}/T]^2}{12}
    - \frac{\pi [g\lambda_{mn}/T]^3}{12} + \frac{[g\lambda_{mn}/T]^4}{48} \bigg\}
    \nonumber\\
    &&\quad\qquad\quad +  \frac{\pi |g\lambda_{mn}/T|^3}{12} \bigg]
  \;,
\end{eqnarray}
where $[x] := x \bmod 2\pi$, the expression in the curly brackets is the GPYW potential in which
the contribution from the zero mode is included and the last term is the zero mode subtraction where
it is taking an absolute value, not modulo $2\pi$.
The details of the derivation are in Appendix~\ref{app:bosonDet}.

This is not quite the whole story.
The logarithm in the first line of Equation~(\ref{eqn:cV}) has the argument
\begin{equation}
  (p_i/T)^2 + (2\pi l - g\lambda_{mn}/T)^2
  \quad\text{where}\quad
  l \in \mathbb{Z}
  \;.
\end{equation}
In the three dimensional point of view, the second term is the effective mass of the Matsubara modes
in units of temperature.
Say, when $g\lambda_{mn}/T = 2\pi$,
the mode $l=1$ becomes massless and the zero mode becomes heavy with the mass $2\pi$;
{\it the lightest mode is not necessarily the zero mode}.
The zero mode becoming heavier than other modes
is not permissible as the separation of scales -- the fundamental premise of effective theories --
is completely ruined.%
\footnote{One could think of constructing an effective theory of the lightest modes, instead of the zero modes, but this theory
  in general would not be time-independent, three dimensional.}
Thus the following constraint is necessary:
\begin{equation}\label{eqn:range}
  |g\lambda_{mn}/T| \ll \pi
  \;.
\end{equation}
Now, the polynomial in the curly brackets of Equation~(\ref{eqn:cV}) is
even with respect to the parameter $g\lambda_{mn}/T$ -- one of the remarkable properties of the polynomial --
as required by the discrete symmetry of the theory.
Therefore, within the range $|g\lambda_{mn}/T|<2\pi$, the cubic term cancels against the zero mode subtraction term
yielding
\begin{eqnarray}
  V_\text{3D} = \frac{2T^3}{\pi^2}{\sum_{m,n}}'\bigg\{
    -\frac{\pi^4}{90} + \frac{\pi^2 (g\lambda_{mn}/T)^2}{12}
    + \frac{(g\lambda_{mn}/T)^4}{48} \bigg\}
\end{eqnarray}
with the condition (\ref{eqn:range}).

To recast this expression in terms of the field, recall that the adjoint representation can be
defined as $(\hat{A}_0^\text{adj})^{ab}:=2\tr(t^a[\hat{A}_0,t^b])$ and then it is easy to show that
$\tr[(\hat{A}_0^\text{adj})^{2l}] = {\sum_{m,n}}'(\lambda_{mn})^{2l}$.
Hence we obtain the final expression
\begin{eqnarray}\label{eqn:V3D}
  V_\text{3D} &=& \tr\bigg[\, \frac{1}{2}\big(\frac{1}{3}g^2T\big)\big(\hat{A}_0^\text{adj}\big)^2
    + \frac{g^4}{4!\pi^2T}\big(\hat{A}_0^\text{adj}\big)^4 \bigg]
  \nonumber\\
  &=& \tr\bigg[\, \frac{1}{2}\big(\frac{1}{3}g_3^2T\big)\big(A_0^\text{adj}\big)^2
    + \frac{g_3^4}{4!\pi^2T}\big(A_0^\text{adj}\big)^4 \bigg]
  \nonumber\\
  &=& \frac{N}{3}g_3^2T\tr[A_0^2]
  + \frac{g_3^4}{12\pi^2T}\bigg\{ 3\big(\tr[A_0^2]\big)^2 + N\tr[A_0^4] \bigg\}
  \;,
\end{eqnarray}
where the field is now understood to be space-dependent $q$-number operator,
I have dropped the identity operator with the coefficient $-(N^2-1)(\pi^2/45)T^3$,
I used the relations $g_3=g\sqrt{T}$ and $A_0^\text{adj}=\hat{A}_0^\text{adj}/\sqrt{T}$,
and the last line is converted to the fields in the defining representation.
See also Nadkarni~\cite{Nadkarni:1988fh} and Landsman~\cite{Landsman:1989be} for the same result.

%%%%%%%%%%%%%%%%%%%%%%%%%%%%%%%%%%%%%%%%%%%%%%%%%%%%%%%%%%%%%%%%%%%%%
\section{The GPYW Potential: A Critique}\label{sec:critique}
%%%%%%%%%%%%%%%%%%%%%%%%%%%%%%%%%%%%%%%%%%%%%%%%%%%%%%%%%%%%%%%%%%%%%
Let me scrutinize the potential.
It was derived in Equation~(\ref{eqn:cV}) and is reproduced here:
\begin{eqnarray}
  V_\text{GPYW} = \frac{2T^4}{\pi^2}{\sum_{m,n}}'\bigg\{ -\frac{\pi^4}{90} + \frac{\pi^2 [g\lambda_{mn}/T]^2}{12}
    - \frac{\pi [g\lambda_{mn}/T]^3}{12} + \frac{[g\lambda_{mn}/T]^4}{48} \bigg\}
  \;,
\end{eqnarray}
where $[x]:=x\bmod 2\pi$, $\lambda_{mn}:=\lambda_m-\lambda_n$, $\lambda_n$ being the eigenvalues of
the constant background $A_0$ field and I supplied an extra overall factor of $T$ for 4D interpretation.
\\
\\%
{\bf Why Does It Truncate at Quartic Order?}

First of all, this is an extremely peculiar result in that it terminates at the quartic term.
One usually expects a one-loop potential to contain infinitely many terms.
Similar computations of scalar potentials in several theories ($\phi^4$, abelian-Higgs,
$\mathcal{N}=4$ super Yang-Mills) show that they all generate infinite terms.
But in our case here, a remarkable identity brings infinite terms
into a finite polynomial [see Equation~(\ref{eqn:thesum}) in Appendix~\ref{app:bosonDet}].

Instead of using this identity, let us expand the cosine;
\begin{equation}\label{eqn:cosineexpansion}
  \sum_{k=1}^{\infty} \frac{\cos(kx)}{k^{d+1}} = \sum_{n=0}^{\infty}\frac{(-1)^n}{(2n)!}\zeta(d+1-2n)x^{2n}
  = \zeta(4) - \frac{1}{2}\zeta(2)x^2 + \frac{1}{4!}\zeta(0)x^4
  \;,
\end{equation}
where $\zeta(s):=\sum_{k=1}^\infty1/k^s$ and the infrared divergences are dimensionally regularized as $d\to 3$.
This obviously is a wrong IR treatment but it simply eliminates all infrared divergences of the zero modes,
leaving only the finite contributions from the heavy modes.
Notice that the right-hand side of the equation truncates at the quartic order because $\zeta(-2n)=0$ for $n\geq1$;
so there is no heavy mode contribution beyond this order.
This series corresponds to the diagram expansion with respect to the background external fields;
for example, the quartic term represents the one-loop diagrams with four external legs carrying zero momenta.
Thus the one-loop diagrams with more than four legs are not just ultraviolet finite
but the heavy modes are somehow conspiring to cancel among themselves,
leaving only the infrared divergences of the zero mode.
(One can easily check this directly by examining the fermionic SUSY counterparts of the diagrams.)
The cancellation is remarkable and this does not, or rather, should not happen without a good reason.
Unlike other theories, what is special about the present theory is that the $A_0$ background behaves
exactly like an imaginary chemical potential (of a phantom charge), making the GPYW potential periodic.
This periodicity enforced in part by the zero modes is somehow forcing the heavy mode
contributions to vanish.
But it remains to be seen exactly how this happens.

Bear in mind that Equation~(\ref{eqn:cosineexpansion}) actually contains very bad infrared-divergent diagrams
of the zero modes which were unjustifiably regularized on the right-had side.
\\
\\%
{\bf $V_\text{GPYW}$ is not $V_\text{3D}$}

Sometimes in the literature, the GPYW potential is claimed to be the potential appearing in the effective
action of the zero modes. As I have demonstrated in the previous section, this is incorrect.
The GPYW potential includes the contributions from the zero modes, meaning the zero modes
are also integrated out.
As such, it is just a number or infinity, not a polynomial of operators.

The zero-mode contribution is non-analytic in $g^2$, as can be seen in the last term of Equation~(\ref{eqn:cV}).
The non-analyticity is the telltale sign of the soft contribution and cannot be generated by perturbation in $g$
at any finite order; it requires summation of infinitely many diagrams.
In the diagram expansion represented in Equation~(\ref{eqn:cosineexpansion}), the infinitely many
infrared-divergent diagrams, which are hidden by dimensional regularization,
sum up to yield the missing $g^3$ non-analytic term.
(Take notice of this very disturbing fact.)
With this in mind, one can see that the zero mode effects are interwoven into the GPYW potential.
For example, if $0<x<2\pi$, then $[-x]^4 = (2\pi-x)^4$;
this contains non-analytic terms proportional to $x$ and $x^3$ (the linear term cancels against the ones from the other terms).
We thus clearly see that the zero mode is integrated out in obtaining the GPYW potential;
again, the GPYW potential is not the 3D effective potential.
\\
\\%
{\bf Not Necessarily MQCD\pmb{$_3$}}
  
Assuming that EQCD$_3$ can be perturbatively constructed,
the correct one-loop 3D potential for an arbitrarily high temperature
is Equation~(\ref{eqn:V3D}) with the condition (\ref{eqn:range}).
This potential clearly has the minimum
at the origin, so at the ``classical'' level, we have $\langle A_0 \rangle = 0$.
This is a classical result in a sense that the zero-mode loops have been deliberately put off.
It is also clear that the potential is harmonic in the neighborhood of
the groundstate and the mass is given by the quadratic term of $V_\text{3D}$.
Now, it is often argued that this $A_0$ mass provides another scale in the system and one can further
``integrate out $A_0$'' to construct a lower scale effective theory.
As I have explained in the previous section, ``integrate out'' in this context should be considered as
a nomenclature for constructing a candidate effective theory by the matching method.
Because the $A_0$ field is out, a candidate is a three dimensional pure
$SU(N)$ Yang-Mills theory of $A_i$ (MQCD$_3$), not restricted to renormalizable operators.
Is this candidate really a plausible effective theory? Not necessarily so.
I have emphasized that the quantum effects of the zero modes have not been taken into account in the potential.
If such effects, especially in the infrared sector, modify the thermal groundstate so that
the $A_0$ scalar develops a nonzero expectation value, then many components of $A_i$ field
acquire masses and the lower scale theory would not look anything like $SU(N)$ Yang-Mills;
it would be a bad candidate for the low energy description.
We must have correct information about the structure of the groundstate to guess
an appropriate candidate.
So why do we believe in MQCD$_3$? Surely, it's the GPYW potential.
\\
\\%
{\bf $\pmb{\langle A_0 \rangle}$ Is Physical in the Static Gauge.}
  
The GPYW potential is a 1PI effective potential that includes the zero-mode quantum effects at one-loop level.
GPYW say that the infrared effects do not significantly modify the groundstate
and the minimum remains at the origin;
that is $\langle A_0\rangle = 0$, making MQCD$_3$ plausible.
The thesis of this work is to argue against this.
But even before that, we must call the physical relevance of this observation into question.

A 1PI effective potential has physical meaning only in its value at the minimum (the energy density).
In general, the location of the minimum is gauge dependent
and has no physical bearing \cite{Jackiw:1974cv,Andreassen:2014eha}.
It should rather be surprising that the GPYW potential is gauge independent
(if one overlooks the underlying potential supersymmetry), but this is not true at higher loops.
Therefore, a gauge must be judiciously chosen so that the resulting gauge-dependent $\langle A_0\rangle$
is useful in extracting relevant physical information.

GPY and the two-loop computations mentioned in the Introduction all employ the background field method with
the background field gauge (an $R_\xi$-like gauge with a parameter $\xi$).
Let us briefly review this procedure.
To make the computations feasible, the background $A_0$ field is always chosen to be constant, diagonal;
this choice must be justified.
A formal 4D gauge symmetry involving transformations of background fields
is left intact in the background field gauge \cite{Abbott:1981ke,Weinberg:1996kr}.
Thus we have the fixed true gauge symmetry and the unfixed formal background gauge symmetry.
The main point of the background field method is to keep the latter symmetry unbroken so that the result
of the integration over the dynamical fields is highly constrained in its form.
But if one chooses, the background gauge symmetry can also be fixed.
One can impose the static gauge and then diagonalize the $A_0$ background field using the remaining
3D spatial background gauge symmetry (recall that the static $A_0$ transforms homogeneously under the 3D transformation).
This diagonal background may be set also to spatial constant for a non-derivative potential.
Thus, the particular form of the background $A_0$ is a formal gauge choice for fixing the background gauge symmetry.
We see that in this procedure, two gauges are involved: one is the background field gauge on the fluctuating fields
and the other is a static gauge on the background fields.
The resulting two-loop potential is dependent on these two gauges and it was observed that the potential
and its location of minimum explicitly depend on the gauge fixing parameter $\xi $ of the background field gauge.
This is only natural, because the eigenvalues of $A_0$ are gauge covariant and their expectation values generally
depend on how the gauge is fixed.
Then $\langle A_0\rangle$ can be zero or nonzero depending on the choice of $\xi$;
this clearly is not physical.

The gauge dependence of $\langle A_0 \rangle$ of course is not a failure of the theory.
The question becomes if one can extract useful physical consequences from this observation.
The arbitrary $\xi$-dependence of $\langle A_0 \rangle$ shows that the effective potential
in the background field gauge
yields no direct physical information about the groundstate other than the energy density.
Put it differently, the background field gauge and other similar gauges are not suitable in drawing
useful conclusions about $\langle A_0 \rangle$ from the effective potential.%
\footnote{See also Arnold \cite{Arnold:1992fb} which criticizes the use of the background field gauge
in computations of effective potential.
He argues that varying the background field to determine the minimum of the potential
does not make sense as it implies varying the gauge itself in the process.}

Instead of the two gauges, wouldn't it be simpler
to just gauge the dynamical $A_0$ field to time-independent diagonal form and then introduce
a background for this diagonal field?
This, in fact, is the gauge adopted by Weiss \cite{Weiss:1980rj} and probably only by him.
The result agrees with GPY's because the computation is at the gauge-independent one-loop level.
Let us see if Weiss's way allows for the extraction of physical information at higher loop orders.
The gauge is a static gauge.
As just mentioned, the $A_0$ field with $\partial_0A_0=0$ transforms homogeneously
under the remaining spatial 3D gauge transformations.
Therefore, the eigenvalues of $A_0$ are gauge invariant under the 3D gauge symmetry.
So a slightly more general procedure than Weiss's can be adopted:
instead of fixing the 3D symmetry by diagonalizing (dynamical) $A_0$,
one can impose a static gauge in which
the 3D gauge fixing is left in a general form with a parameter $\xi_\text{3D}$ (see Nadkarni \cite{Nadkarni:1982kb}
or Appendix~\ref{app:BRST} for the actual gauge fixing term) and then introduce a constant, diagonal
background $A_0$ field, representing the 3D gauge invariant eigenvalues.
We know that the resulting potential will have its minimum at a $\xi_\text{3D}$-independent location; that is,
$\langle A_0 \rangle$ is 3D gauge independent.

So far so good, but the expectation value still is specific to the static gauge, $\partial_0A_0=0$.
I am going to explain that the eigenvalues of $A_0$ under this gauge are physical.
Consider the temporal Wilson loop, the Polyakov loop:
\begin{eqnarray}\label{eqn:defPloop}
  L(\vec{x}) := \frac{1}{N}\tr \mathcal{P}\exp\bigg[\, ig\int_0^{1/T}\!\!dt\, A_0(t,\vec{x}) \bigg]
  \;,
\end{eqnarray}
where $\mathcal{P}$ indicates path ordering. This is a gauge invariant operator.
In this general form, the Polyakov loop and $A_0(t,\vec{x})$ are not in one-to-one correspondence,
but in the static gauge, they are:
\begin{eqnarray}\label{eqn:staticPloop}
  L(\vec{x}) = \frac{1}{N}\tr \exp\big[\, ig A_0(\vec{x})/T \big]
  = \frac{1}{N} \sum_{n=1}^{N}e^{ig\lambda_n(\vec{x})/T}
  \;,
\end{eqnarray}
where the eigenvalues of $A_0$ are denoted as $\lambda_n$ which sum up to zero.
This shows that the eigenvalues of $A_0$, in the static gauge, are directly related to a fully
gauge invariant quantity.

The physical nature of the $A_0$ expectation value is most explicit when $N=2$.
Marhauser and Pawlowski \cite{Marhauser:2008fz} showed that,
instead of $\langle L \rangle$, the expectation value of $A_0$ in the static gauge can be used as an order parameter
of the $\mathbb{Z}_N$ symmetry;
this is not obvious because $\langle L[A_0]\rangle \neq L[\langle A_0\rangle ]$.
They showed that for the positive eigenvalue of $N=2$,
\begin{eqnarray}\label{eqn:pawlowski}
  \langle L(\vec{x}) \rangle = 0   \; \Leftrightarrow \; g\langle\lambda(\vec{x})\rangle/T = \pi/2
  \qquad\text{and}\qquad
  \langle L(\vec{x})\rangle \neq 0 \; \Leftrightarrow \; g\langle\lambda(\vec{x})\rangle/T < \pi/2
  \;,
\end{eqnarray}
The eigenvalues in the static gauge are as real as
the deconfinement phase transition, at least for $N=2$.%
\footnote{Marhauser and Pawlowski assume diagonal form of $A_0$, on top of the static gauge,
  but this is not necessary for the relations in Equation~(\ref{eqn:pawlowski}).
  They claim, without proof, that the relations can be generalized to other $N$, but I fail to see how.}

As mentioned above, the static gauge does not affect the three dimensional spatial gauge symmetry
and the eigenvalues of $A_0$ are invariants of this 3D gauge symmetry.
The expectation values of the eigenvalues, then, have the physical consequence on how the 3D gauge
symmetry is realized by the groundstate and tell
what the useful shape of 3D effective theory should be.
(The eigenvalues are gauge invariant but $A_0$ is not, so nonzero average eigenvalues
can still spontaneously break the symmetry.)
Concretely, if they are zero, then MQCD$_3$ is a reasonable candidate, while if not zero,
then the theory with broken symmetry [e.g. $U(1)^{N-1}$] should be plausible.
In other words, the static gauge captures the zero mode infrared groundstate structure of the theory
through the $A_0$ field.

Thus one can directly obtain physical information from a higher loop effective potential
provided that the static gauge is adopted, though this gauge may not be a convenient choice for the actual computation.
This, however, is a futile effort because the potential is plagued by infrared divergences,
and non-perturbative effects should become dominant.
\\
\\%
{\bf The Infrared Trouble}

In order to understand the trouble, it is useful to examine something very familiar: the renowned result of
Coleman and Weinberg~\cite{Coleman:1973jx}.
The theory is four dimensional massless scalar electrodynamics at zero temperature
and the one-loop 1PI effective potential of the scalar field is computed.
Jackiw in Reference~\cite{Jackiw:1974cv} introduced the background field method for this.
In this method, the scalar field $\phi$ is shifted by a constant background: $\phi + \hat{\phi}$.
Upon the shift, the fields acquire mass terms that depend on the background $\hat{\phi}$.
The crucial fact for us here is that the mass terms are effectively acting as infrared regulators
in the computation.
As such, if $\hat{\phi}$ vanishes, the theory suffers from infrared divergence and {\it the potential at
  $\hat{\phi}=0$ is not defined}.

It is instructive also to look at this pathology in Coleman and Weinberg's original way.
They sum up infinitely many infrared divergent diagrams, similar to Equation~(\ref{eqn:cosineexpansion}).
Schematically, the sum takes the form
\begin{eqnarray}\label{eqn:badExpansion}
  &&\sum_{n=1}^\infty \int \frac{d^dp}{(2\pi)^d} \frac{1}{n}\bigg( -\frac{x^2}{p^2} \bigg)^n
  \,\to\, \int \frac{d^dp}{(2\pi)^d} \sum_{n=1}^\infty \frac{1}{n}\bigg( -\frac{x^2}{p^2 + \mu^2} \bigg)^n
  \nonumber\\
  &&= - \int \frac{d^dp}{(2\pi)^d} \ln \bigg( 1 + \frac{x^2}{p^2 + \mu^2} \bigg)
  \nonumber\\
  && \,\to\,  - |x|^d \int \frac{d^dk}{(2\pi)^d} \ln \bigg( 1 + \frac{1}{k^2} \bigg)
  \;.
\end{eqnarray}
where $d=4$, $x$ is the coupling constant times $\hat{\phi}$ and $k:=p/|x|$.
Coleman and Weinberg regularize the rampant infrared divergences of the first term
by exchanging the order of the integral and the sum.
To justify this operation, both the integral and the sum must be finite.
The regulator $\mu$, therefore, is introduced in the first line with the condition $\mu^2 > x^2$.
Without this regulator, we get a contradiction where the first term is completely infrared divergent
and the last term is completely infrared finite. (The disturbing fact that infinitely many infrared divergent
diagrams yield a finite term.)
The sum can be carried out in a closed form as in the second line.
Finally in the last line, the IR regulator is {\it analytically continued} to zero.
The last term makes it clear that the infrared contribution is finite and proportional to $x^d$.%
\footnote{Contributions near a UV cutoff is not necessarily proportional to $x^d$
  because the cutoff momentum in $k$ is $x$ dependent.}

Now I claim that the analytic continuation is invalid.
To see this, let us examine the second line of Equation~(\ref{eqn:badExpansion}).
The integral yields the terms including the ones proportional to
\begin{equation}\label{eqn:nonanalytic}
  -\mu^4\ln\frac{\mu^2}{M^2} + (x^2+\mu^2)^2\ln\bigg(\frac{x^2+\mu^2}{M^2}\bigg)
  \;,
\end{equation}
where the number of dimensions is explicitly set to four and $M$ is a renormalization scale.
This is not analytic at $\mu=0$, so the parameter may not be continued to this point.
For $x\neq0$, there exists $\mu\neq 0$ such that $(\mu/M)^2$ and $(\mu/x)^2$ are both negligibly small
and the regulator does not affect the integral.
At the same time,
for $\mu\neq0$, there exists a neighborhood of $x=0$ such that $(\mu/x)^2$ is large, say larger than $1$.
Hence in the neighborhood, $\mu$ cannot be ignored relative to $x$,
significantly affecting the integral with the unphysical regulator.
Thus the exchange of the integral and the sum in the first line of Equation~(\ref{eqn:badExpansion})
is not allowed in the neighborhood; there, the potential is as infrared divergent as the first term of the equation.
The cases lead to the conclusion that the potential is infrared divergent at $x=0$.

After the illegitimate analytic continuation to $\mu=0$,
the infrared divergence, therefore, manifests itself as the non-analyticity of the potential at the origin.
Much with the benefit of hindsight, this can also be seen as follows:
\begin{eqnarray}
   \int \frac{d^dp}{(2\pi)^d} \sum_{n=1}^\infty \frac{1}{n}\bigg( -\frac{x^2}{p^2} \bigg)^n
  &\sim& - \int \frac{d^dp}{(2\pi)^d} \ln (p^2 + x^2) \nonumber\\
   &=& - \int \frac{d^dp}{(2\pi)^d} \big[\ln (p + ix)+\ln (p - ix)\big]
   \;,
\end{eqnarray}
where ``$\sim$'' implies that an $x$-independent term has been dropped, for it does not affect the shape of the potential.
Remembering that $d=4$,
the last line makes it clear that the {\it fourth} and higher derivatives with respect to $x$ at the origin
are infrared divergent and do not exist, as observed by Coleman and Weinberg.
The non-analyticity is an expression of the uncontrolled severe infrared divergence and
the potential is ill-defined at the origin.
This singularity at the origin persists to an arbitrary order of perturbation: the propagators are IR regulated
by the background field and this fails at the origin. See also the appendix of Coleman and Weinberg~\cite{Coleman:1973jx}.

It so happens that Coleman and Weinberg find a groundstate away from the origin
and the infrared divergence discussed here is unphysical and irrelevant.
\\

Let us get back to the GPYW potential.
The theory is different but the computation is very similar in the background field method.
As one can see in Equation~(\ref{eqn:cV}), the background field acts as an infrared regulator
for the zero mode.
For $V_\text{3D}$, this regulator and inappropriate soft contributions are subtracted,
but $V_\text{GPYW}$ contains them; the latter is in trouble at the origin.

In the diagram expansion, the schematic zero-mode contributions are exactly the same as
Equation~(\ref{eqn:badExpansion}), except that $d=3$.
(To check this, use the fermionic SUSY counterparts of the diagrams focusing on a Cartan-colored external field.
Notice that the infrared contribution is proportional to $|x|^3$.)
Then we immediately see that the GPYW potential suffers from the infrared divergence and it appears
as the ill-defined {\it third} and higher derivatives at the origin.
This is consistent with the fact that infrared divergences are exacerbated as the dimension is lowered.%
\footnote{It is not difficult to compute the effective potential of {\it doubly} compactified Yang-Mills theory,
  when the two scalars corresponding to the compactified components of the vector field are oriented in
  the flat directions of the color space.
  The resulting effective potential has the ill-defined {\it second} derivative at the origin which is the minimum.}

And the groundstate of the theory, according to the GPYW potential, sits right on top of the infrared divergence.
This invalidates the potential all together as its groundstate energy density is undefined
and we are unable to draw conclusions about the expectation value of $A_0$.
In one sense, this is the failure of the infrared regularization as the background field did not borne out
to appropriately play that role.
In another, the perturbative zero-mode effect slightly away from the origin
is too weak to compete against the heavy mode contribution
and failing to dislocate the groundstate, but at the origin its uncontrolled infrared divergence
catastrophically destroys the perturbative groundstate.

This is an indication that thermal perturbation around the trivial groundstate
does not make sense.
One could postulate that some non-perturbative effects provide infrared cutoff to make sense out of the perturbation.
But Linde has shown \cite{Linde:1980ts} that such perturbation around the trivial groundstate
breaks down beyond certain orders -- at best.
It is more natural to regard Linde's problem as just another indication of the trouble with
the trivial groundstate.

Given the failure of the one-loop result, it is unlikely
that the two-loop version of the effective potential yields physical, non-vanishing
$A_0$ expectation value.
It is more likely that
the two-loop computation in the static gauge -- if the computation is feasible --
would yield vanishing expectation value at which the potential is infrared divergent.
Notice that all along, including Linde's analysis,
the problem has been the infrared divergences of essentially three dimensional Yang-Mills theory.
The vacuum of the 3D theory confines and is highly non-perturbative; this is the culprit of the divergence
at the origin.
Then no perturbation theory can get around the infrared divergence and it should always yield
divergent groundstate at the origin at any order.

To summarize, the hot vacuum structure is likely dominated by the non-perturbative effects of the zero mode and
the $A_0$ field would acquire nontrivial expectation values.
Further investigation along this line clearly requires non-perturbative means, like lattice.

%%%%%%%%%%%%%%%%%%%%%%%%%%%%%%%%%%%%%%%%%%%%%%%%%%%%%%%%%%%%%%%%%%%%%
\section{Lattice Simulations}\label{sec:lattice}
%%%%%%%%%%%%%%%%%%%%%%%%%%%%%%%%%%%%%%%%%%%%%%%%%%%%%%%%%%%%%%%%%%%%%
In this section, I am going to directly obtain the values of $\langle A_0\rangle$ on the lattice,
focusing on the gauge group $SU(2)$.
Sections \ref{subsec:obs} and \ref{subsec:takeaways}, a short concluding summary,
are accessible to lattice non-experts.
The technical simulation details are described in Appendix~\ref{app:details}.

%%%%%%%%%%%%%%%%%%%%%%%%%%%%%%%%%%%%%%%%%%%%%%%%%%%%%%%%%%%%%%%%%%%%%
\subsection{The Observable}\label{subsec:obs}
Let us first establish the observable. The operator of interest is the aforementioned Polyakov loop
reproduced here for $N=2$:
\begin{eqnarray}\label{eqn:su2PLoop}
  L(\vec{x}) := \frac{1}{2}\tr \mathcal{P}\exp\bigg[\, ig\int_0^{1/T}\!\!dt\, A_0(t,\vec{x}) \bigg]
  \;,
\end{eqnarray}
In the lattice setup, this is given by the traced product of temporal links ordered in the time direction
at a spatial lattice site $\vec{x}$ and it winds once around the periodically identified time direction.

In the foregoing sections, the importance of the static gauge was emphasized repeatedly.
This gauge, therefore, is adopted in this section.
Under the static gauge, we have a simple relation
\begin{eqnarray}\label{eqn:LA0}
  L(\vec{x}) = \cos\big(gA_0(\vec{x})/T\big)
  \quad\text{with}\quad
  A_0(\vec{x}) := \lambda_1(\vec{x}) = - \lambda_2(\vec{x})
  \;.
\end{eqnarray}
The reader is warned for the abuse of notation: the eigenvalues of the $A_0$ matrix
are now denoted as $\pm A_0$, instead of $\lambda_n$.
The new notation is intuitively more appealing.
In the following, I assume positive $A_0$.
The relation directly leads to the target observable
\begin{eqnarray}\label{eqn:A0}
  g\langle A_0(\vec{x}) \rangle/T = \langle \arccos L(\vec{x}) \rangle
  \;.
\end{eqnarray}
This relation yields $\langle A_0\rangle$ by measuring $L$.
Because $L$ is gauge invariant, I can measure $L$ in any gauge; then Equation~(\ref{eqn:A0})
provides the $A_0$ expectation value in the static gauge.
In fact, I am not going to choose a gauge to measure $L$, for given sufficient numerical precision
(such as {\tt float} or {\tt double}) the probability of sampling more than one configuration from
a single gauge orbit is practically zero.

Finally, note that $A_0$ at one spatial position is as good as another.
Hence I build a Euclidean invariant operator by averaging it over the three-space
and then take expectation value of it:
\begin{eqnarray}\label{eqn:A0measure}
  g\langle \bar{A}_0 \rangle/T := \bigg\langle \frac{1}{N_s^3}\sum_{\vec{x}} \arccos L(\vec{x}) \bigg\rangle
  \;,
\end{eqnarray}
where $N_s$ is the spatial extent of the lattice in units of lattice spacing.

%%%%%%%%%%%%%%%%%%%%%%%%%%%%%%%%%%%%%%%%%%%%%%%%%%%%%%%%%%%%%%%%%%%%%
\subsection{Simulation Strategy}\label{sec:simStrategy}
The standard Wilson action is employed in the simulations with the lattice coupling $\beta_\ell$.
The temperature of the system is given by $T=1/(aN_t)$ where $N_t$ is the temporal extent of the lattice
in units of the lattice spacing $a$ which, in turn, is a function of $\beta_\ell$.

I choose to measure the temperature in units of $T^c$, the deconfinement
transition temperature, so that $T/T^c = a(\beta_\ell^c)N_t^c/a(\beta_\ell)N_t$,
where $\beta_\ell^c$ is the critical coupling of the transition when the temporal extent
of the lattice is $N_t^c$.
The problem is that the exact form of the function $a(\beta_\ell)$ is not known.
Thus, I choose to vary temperature by changing $N_t$ at a fixed $\beta_\ell=\beta_\ell^c$,
leaving us the form $T/T^c = N_t^c/N_t$.
This is superficially independent of the function $a(\beta_\ell)$ but as we will see shortly, the required
set of values $(\beta_\ell^c,N_t^c)$ are related through the function.
The reason why the temperature is varied this way is that even without knowing $N_t^c$, the ratios
of the temperatures are exact; for example, if I halve $N_t$, the temperature is exactly doubled.
The price I have to pay is the limited number of data points, as $N_t$ can take only relatively small even integers.
(It has to be even because of my thermalization algorithm.)
We will, however, find sufficient data points to convince us the shape of the data in
a given range of temperature.
\\

In the discussion following this subsection, it will be useful to have a reasonable form of $a(\beta_\ell)$
so that I know a fair value of $N_t^c$ for a given $\beta_\ell^c$.
The parameter $N_t^c$ can be written as
\begin{equation}\label{eqn:Ntc}
  N_t^c = \frac{1}{aT^c} = \frac{(T^c/\Lambda)^{-1}}{a\Lambda(\beta_\ell^c)}
  \;,
\end{equation}
where $\Lambda$ is a certain scale, usually set to the square root of the zero temperature string tension $\sqrt{\sigma}$
or the lattice scale $\Lambda_L$.
For the function $a\Lambda(\beta_\ell)$, consider a large $\beta_\ell$ expansion
\begin{eqnarray}\label{eqn:block}
  2\ln\big(a\Lambda\big) = -\frac{6\pi^2}{11}\beta_\ell + \frac{102}{121}\ln\bigg(\frac{6\pi^2}{11}\beta_\ell\bigg)
  + c_1 + \frac{c_2}{\beta_\ell^n}
  \;.
\end{eqnarray}
The first two terms are the two-loop perturbation result and the constants $c_{1,2}$ of 
the higher order terms are subject to data fitting.
The parameter $n$ is an integer for which Block {\it et al.} \cite{Bloch:2003sk} suggest $n=1$
while Engels {\it et al.} \cite{Engels:1994xj} $n=3$.
Notice that the factor $T^c/\Lambda$ in Equation~(\ref{eqn:Ntc}) is a constant so this can be absorbed into
the parameter $c_1$ in Equation~(\ref{eqn:block}).
Hence the expression
\begin{equation}\label{eqn:fit}
  -2\ln N_t^c + \frac{6\pi^2}{11}\beta_\ell^c - \frac{102}{121}\ln\bigg(\frac{6\pi^2}{11}\beta_\ell^c\bigg)
  = c_1 + \frac{c_2}{(\beta_\ell^c)^n}
\end{equation}
is fit to data; this is an elementary linear fit problem.
The data $(\beta_\ell^c,N_t^c)$ determined directly on the lattice are collected in Reference~\cite{Fingberg:1992ju}
and I use the four largest data points (see Appendix~\ref{app:details} for why I eschew smaller $N_t^c$):
\begin{table}[h]
  \centering
  \begin{tabular}{c||c|c|c|c}
    $N_t^c$ & $5$ & $6$ & $8$ & $16$ \\    
    \hline
    $\beta_\ell^c$ & $2.3726(45)$ & $2.4265(30)$ & $2.5115(40)$ & $2.7395(100)$ \\
    \hline
    $\delta N_t^c$ & $0.076063$ & $0.059431$ & $0.10145$ & $0.48479$
  \end{tabular}
\end{table}
\\
The top two rows are the data from the reference and I now explain the third row.
The data shown is a little inconvenient because the input is $N_t^c$ and the output is $\beta_\ell^c$ whereas I have $N_t^c$
as a function of $\beta_\ell$.
In order to invert the relation of the data, following procedure is employed.
First, the fit is tentatively done for the data ignoring the errors in $\beta_\ell$ and also setting all the variances to unity.
Then this curve is used to estimate the propagation of the error from $\beta_\ell$ to $N_t^c$.
Finally, treat the $\beta_\ell$s as exact input values at which the corresponding $N_t^c$s are given with the errors,
as quoted in the third row of the table (for the choice $n=3$).
They are used as the weights of the least square fit.

Now we are ready to fit Equation~(\ref{eqn:fit}) to the data.
The fit for $n=1$ and $3$ are both very good with the reduced $\chi^2$s $0.65$ and $0.27$, respectively.
(In fact, the reduced $\chi^2$s are a bit too good and show that the errors are overestimated to some extent.)
The latter is somewhat better and it turns out that the errors propagated to $N_t^c$ through Equation~(\ref{eqn:fit})
are smaller for $n=3$.
Hence I will present the case $n=3$ in what follows.%
\footnote{The parameter $N_t^c$ computed for $n=1$ and $3$ have overlapping error bars for the range of
  $\beta_\ell$ that I have used, so the choice is not mutually exclusive.}
The fit result is
\begin{equation}\label{eqn:fitsoln}
  n=3
  \text{:}\quad
  \left(\begin{array}{c}
    c_1 \\
    c_2
  \end{array}\right)
  =
  \left(\begin{array}{c}
    5.9858 \\
    18.916
  \end{array}\right)
  \quad\text{with}\quad
  \pmb\epsilon = 
  \left(\begin{array}{cc}
    0.0063746 & -0.093041 \\
    -0.093041 & 1.3688
  \end{array}\right)
  \;,
\end{equation}
where I have included the $\hat\chi^2$-improved error matrix $\pmb\epsilon$ whose variances and
covariance are used in subsequent error propagations.%
\footnote{A $\hat\chi^2$-improved error is computed via variances that are the products of
  the original variances and the reduced $\chi^2$ that results from the fit.}

Given this result and Equations~(\ref{eqn:fit}), I can obtain $N_t^c$ for an arbitrary $\beta_\ell^c$.
Table~\ref{tbl:bN} shows the parameters I use in the next subsection.
\begin{table}[h]
  \centering
  \begin{tabular}{c||c|c|c|c}
    $\beta_\ell^c$ & $2.6414184$ & $2.7348566$ & $2.8087297$ & $2.8699112$ \\    
    \hline
    $N_t^c$ & $12.00(44)$ & $16.00(57)$ & $20.000(705)$ & $24.00(84)$ \\
  \end{tabular}
  \caption{\footnotesize{The values of $(\beta_\ell^c,N_t^c)$ used in the simulations.
      They are obtained from Equations~(\ref{eqn:fit},\ref{eqn:fitsoln}).}}
  \label{tbl:bN}
\end{table}

Summary of the strategy: I thermalize a lattice of temporal extent $N_t$ at the coupling $\beta_\ell^c$ from the table;
then measure the Polyakov loops and call them the values at the temperature
$T/T^c=N_t^c/N_t \pm \delta N_t^c/N_t$. This procedure is repeated for different values of $N_t$.

%%%%%%%%%%%%%%%%%%%%%%%%%%%%%%%%%%%%%%%%%%%%%%%%%%%%%%%%%%%%%%%%%%%%%
\subsection{Renormalization of the Polyakov Loop}\label{subsec:ploops}
\begin{figure}[t]
  \centering
  \includegraphics{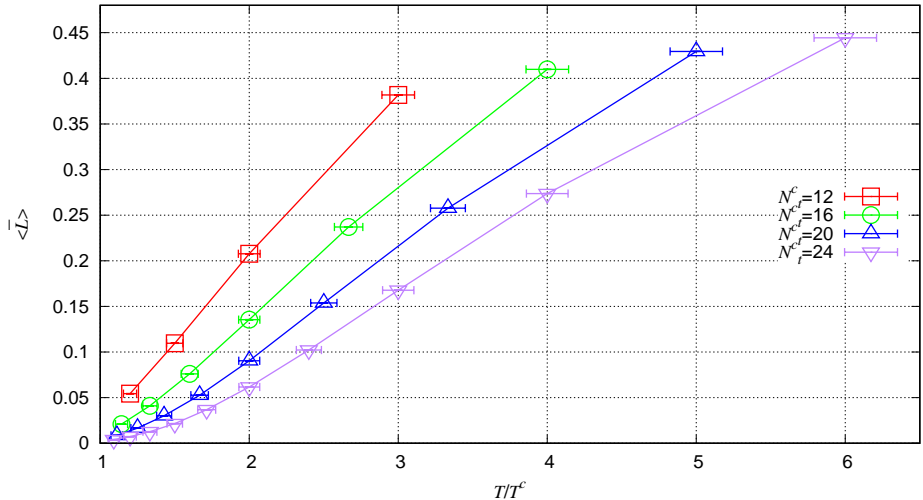}
  \caption{\footnotesize{The space-averaged expectation values of the Polyakov loops for different $(\beta_\ell^c,N_t^c)$.
      See Table~\ref{tbl:bN} for the values of corresponding $\beta_\ell^c$.
      The simulation statistical errors (vertical error bars) and the errors in temperatures (horizontal error bars)
      are both shown but the former is too small to be discernible.
      The values of $N_t$ are even integers in the range [4,$N_t^c-2$], whereas the spatial extent is set at either $64$ or $96$
      so that $4N_t<N_s$.
      The data at $N_t=N_t^c$, that is,  $T/T^c\approx 1$ are not used to avoid possible contamination by the critical slowing down.
      Nor are the data at $N_t=2$ used because of discretization artifacts;
      see Appendix~\ref{app:details}.}}
  \label{fig:bareL}
\end{figure}

Measurement of the Polyakov loop is straightforward.
The space-averaged Polyakov loop is defined as
\begin{eqnarray}\label{eqn:multrenorm}
  \bar{L} := \frac{1}{N_s^3}\sum_{\vec{x}} L(\vec{x})
\end{eqnarray}
and its expectation values are plotted in Figure~{\ref{fig:bareL}}.
The data sets in the figure correspond to the different sets $(\beta_\ell^c,N_t^c)$ shown in Table~\ref{tbl:bN}.
We see that $\langle\bar L\rangle$ does not scale and its continuum limit $\beta_\ell\to\infty$ does not make sense.
This is because of the ultraviolet divergence and a well-defined $\langle\bar L\rangle$ requires renormalization.

Renormalization of the Polyakov loop, or Wilson loops in general, is highly nontrivial because
of its non-local nature.
Polyakov~\cite{Polyakov:1980ca} suggested that the gauge field and the coupling constant are renormalized as usual,
while the linear divergence associated with the mass of a test quark taken along the loop is multiplicatively renormalized.
This was later explicitly shown perturbatively~\cite{Dotsenko:1979wb}.
Gupta {\it et al.}~\cite{Gupta:2007ax} postulated the multiplicative renormalization (non-perturbatively) on the lattice,
which I write as
\begin{eqnarray}\label{eqn:renormL}
  \big\langle L_r(\hat{T})\big\rangle  = Z(\beta_\ell)^{1/\hat{T}}\big\langle L_b(\beta_\ell,\hat{T})\big\rangle
  \;,
\end{eqnarray}
where $L_r$ and $L_b$ are renormalized and bare loops, respectively, $Z(\beta_\ell)$ is the renormalization
factor and I have defined $\hat{T}:=T/T^c$.
(My definition is slightly different from Reference~\cite{Gupta:2007ax}.)
This postulate at the non-perturbative level is reasonable because renormalization is a short distance business
where perturbation theory is valid anyway.
In fact, there is strong evidence that the multiplicative renormalization (\ref{eqn:renormL}) works well
\cite{Gupta:2007ax,Kaczmarek:2002mc}.
One immediate advantage is that $\langle L_r\rangle$ is as good a $\mathbb{Z}_N$
order parameter as the traditional one, $\langle L_b\rangle$,
thanks to the multiplicative nature of the renormalization.
For this reason, I am going to use this renormalization method,
adapted to the strategy described in the previous subsection.

Equation~(\ref{eqn:renormL}) is a statement that the $a(\beta_\ell)$-cutoff dependency of $\big\langle L_b\big\rangle$
is completely removed by the factor $Z(\beta_\ell)$.
Such ``subtraction of infinity'' leaves ambiguity in the finite leftover, i.e., there is a
constant $C$ that can be defined by the relation
\begin{eqnarray}
  Z(\beta_\ell) = C \tilde{Z}(\beta_\ell)
  \;.
\end{eqnarray}
This constant $C$ represents a renormalization scheme.
It may {\it not} be arbitrary but must be chosen such that $|\langle L_r \rangle|\leq 1$ is satisfied for all $\hat T$.
The upper limit exists because the Polyakov loop expectation value is related to the difference of the free energies with
and without a test quark~\cite{Svetitsky:1985ye},
and $|\langle L_r \rangle|>1$ would imply that the hot Yang-Mills vacuum {\it decreases}
energy upon inserting an infinitely heavy quark; such a vacuum is unstable and the theory would cease to exist at
a scheme-dependent arbitrary temperature.
This upper limit can be breached by some schemes and they must be rejected as unphysical.

I choose a scheme
$C = 1/\tilde{Z}(\tilde\beta_\ell)$ for some fixed $\tilde\beta_\ell$, leading to the relation
\begin{eqnarray}\label{eqn:pickScheme}
  \big\langle L_r(\hat{T})\big\rangle = \big\langle L_b(\tilde\beta_\ell,\hat{T})\big\rangle
  \;.
\end{eqnarray}
I emphasize that this equality holds for varying $\hat{T}$, but only for the particular scheme
at the fixed $\tilde\beta_\ell$.
Because $\langle L_b \rangle \leq 1$, this scheme satisfies the condition.
Now, at an arbitrary reference temperature $\hat{T}=\hat{T}_\text{ref}$, we have the relation
\begin{eqnarray}\label{eqn:rescale}
  \big\langle L_r(\hat{T}_\text{ref})\big\rangle = \big\langle L_b(\tilde\beta_\ell,\hat T_\text{ref})\big\rangle
  = \big[ Z(\beta_\ell) \big]^{1/\hat{T}_\text{ref}}\big\langle L_b(\beta_\ell,\hat T_\text{ref})\big\rangle
  \;,
\end{eqnarray}
for an arbitrary $\beta_\ell$. One can solve the equation for $Z(\beta_\ell)$ and plug this into
Equation~(\ref{eqn:renormL}), obtaining
\begin{eqnarray}\label{eqn:renormalized}
  \big\langle L_r(\hat{T})\big\rangle = \big[\big\langle L_b(\tilde\beta_\ell,\hat{T}_\text{ref})\big\rangle /
    \big\langle L_b(\beta_\ell,\hat{T}_\text{ref})\big\rangle \big]^{\hat{T}_\text{ref}/\hat{T}} \big\langle L_b(\beta_\ell,\hat{T})\big\rangle
  \;.
\end{eqnarray}

Let me translate this result into English.
First,  pick a curve in Figure~\ref{fig:bareL} and declare it the renormalized curve in a particular scheme
[Equation~(\ref{eqn:pickScheme})].
Next, pick another curve. Then rescale the value of this second curve at an arbitrary temperature
$\hat T_\text{ref}$ so that it matches the value
of the first curve at the same temperature [Equation~(\ref{eqn:rescale})].
This rescaling factor is the one in the square brackets of Equation~(\ref{eqn:renormalized}).
Finally, rescale the whole of the second curve according to Equation~(\ref{eqn:renormalized});
for instance, the second curve at $\hat T = 2\hat T_\text{ref}$
is rescaled by the square root of the original factor at $\hat T = \hat T_\text{ref}$.
This procedure brings the second curve identical to the first; that is, the second curve is renormalized.
Notice that this simple procedure is enabled by the simulation strategy described in Section~\ref{sec:simStrategy}
which allows $\beta_\ell$ and $\hat T$ to vary independently.

\begin{figure}[t]
  \centering
  \includegraphics{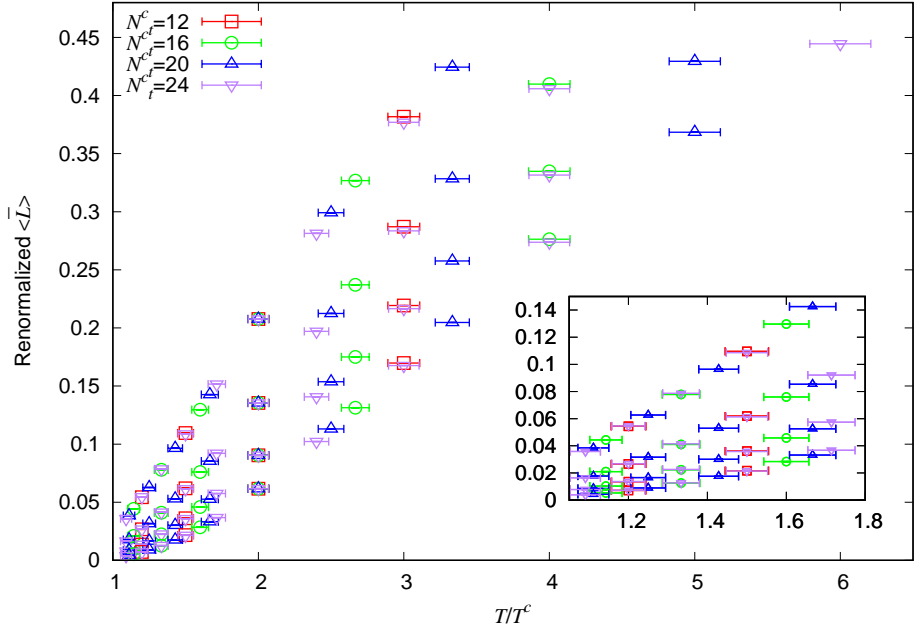}
  \caption{\footnotesize{ The renormalized version of Figure~\ref{fig:bareL}.
      For the leftmost curve, the scheme is selected for $N_t^c=12$ and the data sets of $N_t^c=16,20,24$
      are renormalized in this scheme.
      The other three curves are similarly renormalized for the other schemes.
      The matching point is $\hat T_\text{ref}=2$.}}
  \label{fig:renormL}
\end{figure}

Figure~\ref{fig:renormL} shows the renormalized version of Figure~\ref{fig:bareL}.
(The horizontal error bars are not propagated in the renormalization procedure;
this is because the errors in the temperatures are already overestimated, as I will show shortly.)
Four schemes are shown and we see that the curves are renormalized to a reasonably well-defined single curve
for each scheme.
I am going to quantify how ``reasonable'' this observation is.
The idea is to fit a curve to the $N_t^c=24$ data set and observe the reduced $\chi^2$ of other data sets --
renormalized to the scheme of $N_t^c=24$ -- with respect to the fit curve.
If they are renormalized to a single curve as claimed, then the reduced $\chi^2$ should be about $1$ or less.

I propose a fit curve
\begin{equation}\label{eqn:fitcurve}
  \langle \bar L \rangle = 1 + \big\{a_0 + a_1(\hat T-1) + a_2(\hat T-1)^2 + a_3(\hat T-1)^3\big\}e^{-a_4(\hat T-1)}
  \;,
\end{equation}
where $a_i$ are fit parameters.
At $\hat T=1$, $\langle \bar L \rangle$ should be zero, but because of the errors in temperature, I give it
leeway by introducing the parameter $a_0$; this should come near $-1$ anyway.
At large $\hat T$, exponential approach to the upper limit $\langle \bar L \rangle=1$ is assumed arbitrarily.
This exponent is introduced just to tame the lousy behavior of the polynomial fit, so this is not expected to be accurate
near the upper limit.
Note also that a polynomial fit is generally egregious in extrapolating outside the range of fit data.
This is why I choose to renormalize the data to $N_t^c=24$ whose data has the widest temperature range, so
the renormalized points from the other curves would be interpolating the fit curve.

\begin{figure}[t]
  \centering
  \includegraphics{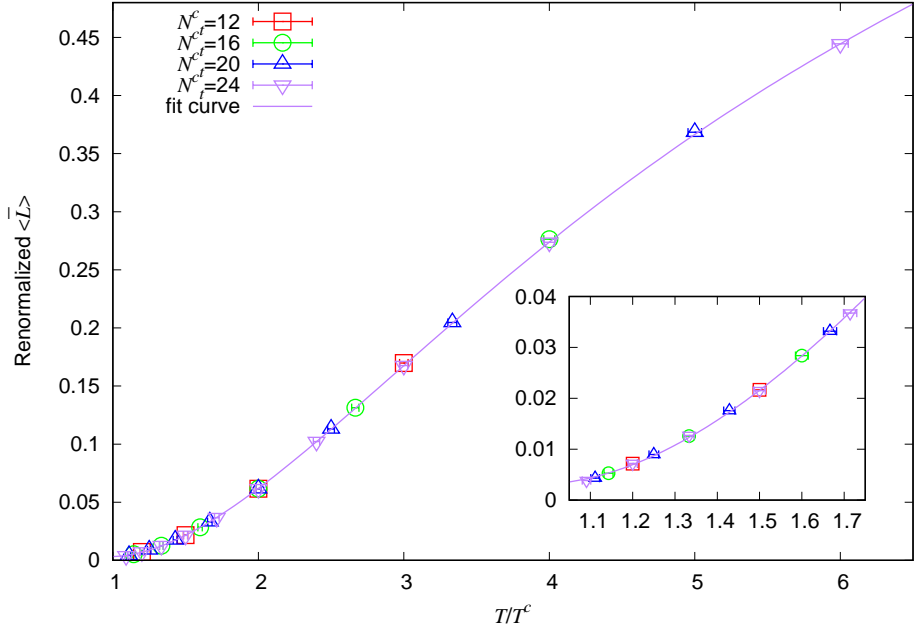}
  \caption{\footnotesize{The data shown in Figure~\ref{fig:bareL} renormalized to the scheme of $N_t^c=24$,
      also the rightmost data set of Figure~\ref{fig:renormL}.
      The solid line is Equation~(\ref{eqn:fitcurve}) fitted to the data set of $N_t^c=24$, the purple-colored data points.
      The fit parameters are $\{a_0,a_1,a_2,a_3,a_4\}=\{-0.9969(4),-0.355(14),0.0133(25),-0.00751(65),0.36(1)\}$.
      The matching point is $\hat T_\text{ref}=2$.
      The error bars in this figure are the $\hat\chi^2$-improved values.}}
  \label{fig:renorm24}
\end{figure}

I have increased the statistics of the simulation so that the statistical errors (vertical error bars) are
well below one percent of the errors in temperature (horizontal error bars); thus, I am going to ignore
the vertical error bars.
The fit function, however, cannot be inverted explicitly, so the least-square fit is carried out
by setting all the variances to unity.
The resulting fit to the data set of $N_t^c=24$ is shown in Figure~\ref{fig:renorm24}.
The fit curve is numerically inverted and compared to the renormalized data points that have the horizontal error bars.
The resulting reduced $\chi^2$s are too small ($\hat\chi^2\sim10^{-2}$), again, indicating that the errors are overestimated.
The $\hat\chi^2$ of the fit to the original data $N_t^c=24$ is $0.069$,
so I am going to multiply all the variances with this value.
By design, the original fit gives $\hat\chi^2=1$ and the idea is to see how the renormalized data with the $\hat\chi^2$-improved
variances would fit to the curve compared to the $N_t^c=24$ data.
(The simulation statistical errors are still about one percent of the new improved errors, so can be safely ignored.)
The result is remarkable: the improved $\hat\chi^2$s of the renormalized data sets $N_t^c=12,16$ and $20$ are $0.21$, $0.12$
and $0.33$, respectively.
(The degrees of freedom are larger for the renormalized data sets because they are not used for the fit,
so the $\hat\chi^2$s can be less than 1.)

This result means three things: {\it i}) the temperature estimated using Equations~(\ref{eqn:fit},\ref{eqn:fitsoln}) is fair,
though the errors are large;
{\it ii}) the fit ansatz (\ref{eqn:fitcurve}) is decent; and {\it iii}) the renormalized data sets are very likely on a single curve.
This {\it iii}) is a strong vindication of the multiplicative renormalization~(\ref{eqn:renormL})
and of the renormalization procedure described above.
This is a verification of the general principle; thus it is highly likely that the renormalization
works outside the range of temperature examined,
even though the observations {\it i}) and {\it ii}) are less likely to extrapolate too far.
\\

I have discussed the renormalization of $\langle L \rangle$ but in the next subsection, we will see that
the computation of $\langle A_0\rangle$ requires the renormalization of $\langle L^{2k+1} \rangle$
for all positive integers $k$.
Equation~(\ref{eqn:renormL}) is no ordinary renormalization and different from the absorption of
the logarithmic divergences
but is the mass renormalization of the test quark \cite{Polyakov:1980ca,Dotsenko:1979wb}.
Consequently, it does not have a corresponding renormalized operator such as
$Z'L$ ($Z':=Z^{1/\hat T}$).%
\footnote{Promoting the renormalization to the operator statement, $Z'L$ instead of $Z'\langle L\rangle$,
  would imply the ``renormalized'' link variables
  taking values outside the $SU(2)$ group elements and has no interpretation as exponential of an anti-Hermitian operator.}
The expectation values $\langle L^{2k+1} \rangle$, therefore, do {\it not} renormalize as $Z'^{2k+1}\langle L^{2k+1} \rangle$.

To see how they renormalize, consider the reduction of $L^{2k+1}$.
Let $\chi_l$ be the $SU(2)$ group character of the irreducible representation $l$.
In particular, I have $2L=\chi_{1/2}$.
Then $L^{2k+1}$ can be expressed as linear combinations of $\chi_l$ where $l$ takes the values of half integers.
The each linear combination always involves a term proportional to $\chi_{1/2}$, which in turn is proportional to $L$;
for instance, $(2L)^3=\chi_{3/2}+2\chi_{1/2}$.
It follows that if $\langle L^{2k+1}\rangle$ renormalize multiplicatively, then they all must renormalize
the same way as $\langle L\rangle$ does: $Z'\langle L^{2k+1}\rangle$.
Now, this is easy to check on the lattice for small values of $k$.
\begin{figure}[t]
  \centering
  \includegraphics{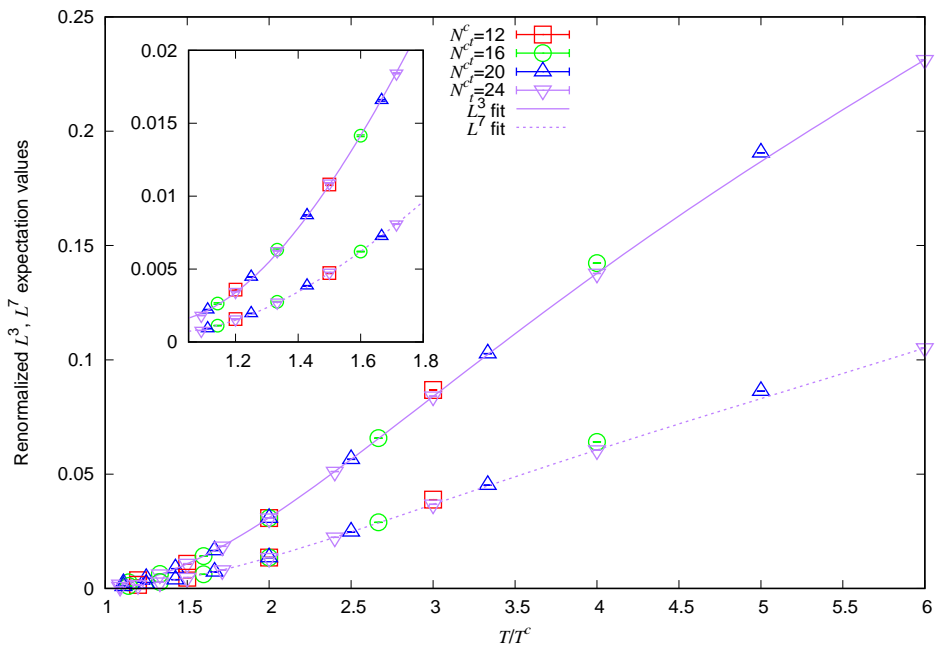}
  \caption{\footnotesize{Renormalized $\langle L^3\rangle$ (top) and $\langle L^7\rangle$ (bottom).
      The $\langle L^{3,7}\rangle$ data for $N_t^c=12$, $16$ and $20$ are renormalized to the scheme of $N_t^c=24$.
      The renormalization factors are identical to the ones used for $\langle L\rangle$ in Figure~\ref{fig:renorm24}.
      The solid and dotted curves are the fits of Equation~(\ref{eqn:fitcurve}) to the data $\langle L^{3,7}\rangle$ of $N_t^c=24$.
      The horizontal error bars are identical to Figure~\ref{fig:renormL} and omitted for this plot.
      The propagation of the horizontal errors to the vertical errors through the renormalization are ignored.      
      Notice the systematic deviations of the data at $T/T^c=3$, $4$ and $5$.
      They are the renormalized data of $N_t^c=12$, $16$ and $20$, respectively, at $N_t=4$.
      They are the discretization artifacts discussed in Appendix~\ref{app:details}.
  }}
  \label{fig:L37}
\end{figure}
Figure~\ref{fig:L37} shows the cases $k=1$ and $3$ with the renormalization factor $Z'$ identical
to the ones used for $\langle L\rangle$, and does demonstrate the expected renormalization property.%
\footnote{The expectation values $\langle L^{2k}\rangle$ do not renormalize as $\langle L\rangle$ does.
It is left for those interested to figure out.}
\\

After a bit of ado, the sole point of this subsection is to demonstrate that the unrenormalized
``bare curves'' in Figure~\ref{fig:bareL} can be interpreted as renormalized
Polyakov-loop expectation values in different schemes.
In the simulation strategy described in the previous subsection, {\it a choice of the set $(\beta_\ell^c,N_t^c)$
corresponds to picking a particular scheme.}
Once a scheme is chosen, all $\langle L^{2k+1}\rangle$ are in the same scheme; this leads to
simple computations of $\langle A_0\rangle$ in the next subsection.

%%%%%%%%%%%%%%%%%%%%%%%%%%%%%%%%%%%%%%%%%%%%%%%%%%%%%%%%%%%%%%%%%%%%%
\subsection{The Result}\label{subsec:result}
\begin{figure}[t]
  \centering
  \includegraphics{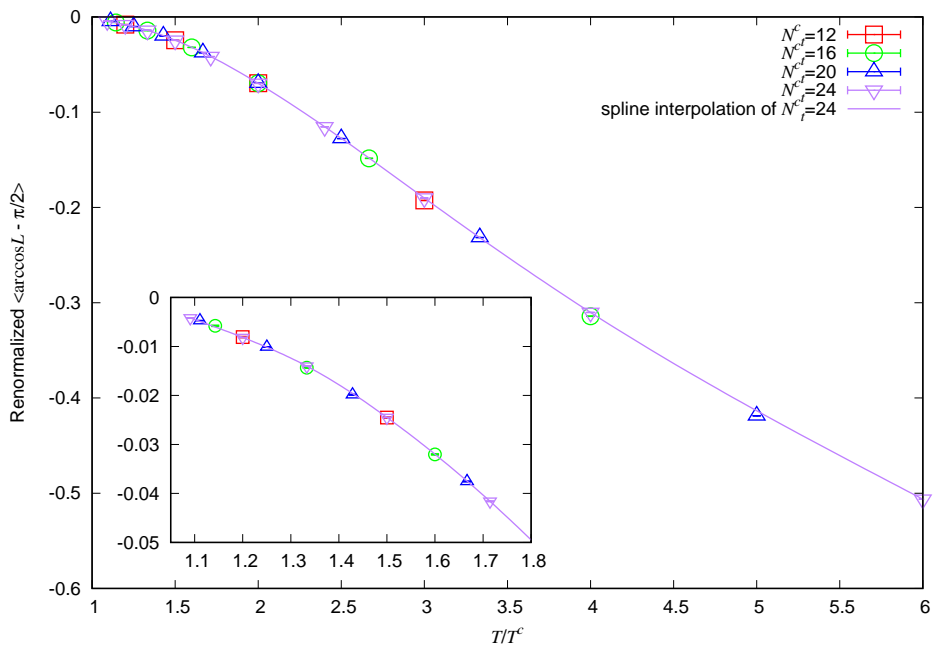}
  \caption{\footnotesize{The renormalized $\langle \arccos L - \pi/2\rangle$.
      The data of the expectation values for $N_t^c=12$, $16$ and $20$ are renormalized to the scheme of $N_t^c=24$.
      The renormalization factors are identical to the ones used for $\langle L\rangle$ in Figure~\ref{fig:renorm24}.
      The solid curve is a spline interpolation of the data for $N_t^c=24$.
      The horizontal error bars are identical with Figure~\ref{fig:renormL} and omitted for this plot.
      The propagation of the horizontal errors to the vertical errors through the renormalization are ignored.
      }}
  \label{fig:a0renorm}
\end{figure}

The expression for the expectation value $\langle A_0\rangle$ in
Equation~(\ref{eqn:A0}) is a short for the formal power series
\begin{equation}\label{eqn:A0expansion}
  g\langle A_0 \rangle/T - \frac{\pi}{2}\langle\mathbf{1}\rangle = \sum_{k=0}^{\infty}a_k\langle L^{2k+1} \rangle
  \;,
\end{equation}
where $\mathbf{1}$ is the identity operator and $a_k$ are expansion coefficients
(see the Table of Integrals \cite{IntegralTable} p.60).
Since the right-hand side renormalizes multiplicatively, so does the left-hand side.
Notice, however, that $\langle A_0 \rangle$ alone does not renormalize properly, i.e., it remains $\beta_\ell$ dependent;
this is because the identity operator is originally $\beta_\ell$ independent so its multiplicative
renormalization introduces the $\beta_\ell$ dependency and this must be canceled by the renormalized $\langle A_0 \rangle$.
Hence what we should really be observing is the combination shown on the left-hand side of Equation~(\ref{eqn:A0expansion})
which properly renormalizes as $\langle L \rangle$ does --
see Figure~\ref{fig:a0renorm} --
and the scheme chosen for $\langle L \rangle$ is shared by the combination.
So as long as I stay in a single scheme $(\beta_\ell^c,N_t^c)$, no {\it re}normalization is necessary.
A straightforward simulation of $\langle \bar A_0\rangle$ through Equation~(\ref{eqn:A0measure})
with {\it bare} $L$ yields a renormalized $\langle \bar A_0\rangle$ in the specific scheme (and in the static gauge).
Also as long as I stick to a single scheme, the errors in temperature are not important and I do not have to worry
about the validity of the fit Equations~(\ref{eqn:fit},\ref{eqn:fitsoln}).
I can always choose to measure the temperature in a certain scale.
Then the ratios of the temperatures are exact
in the strategy described in Section~\ref{sec:simStrategy} and there are no errors in the temperature measured in the scale.
(The error in temperature is important in comparing results of different schemes
because they need to share the same scale.)

\begin{figure}[t]
  \centering
  \includegraphics{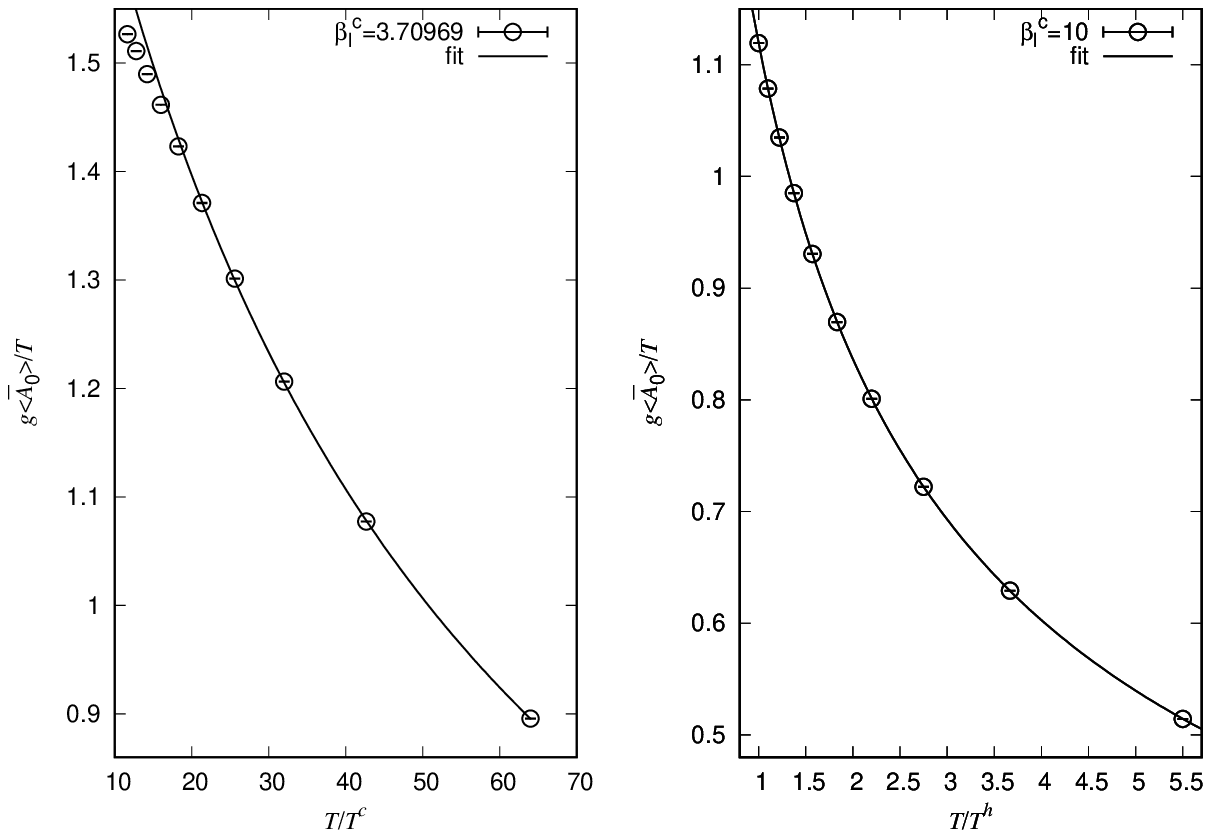}
  \caption{\footnotesize{The plots of $g\langle \bar A_0\rangle/T$ at $\beta_\ell^c=3.70969$ (left) and $10$ (right).
      The former, according to Equations~(\ref{eqn:fit},\ref{eqn:fitsoln}), corresponds to $N_t^c=256$ and this value
      is used for $T^c$ of the horizontal axis.
      Notice, however, that $\beta_\ell^c=3.70969$ is quite a bit of extrapolation from the range of $\beta_\ell^c$ in which
      the fit (\ref{eqn:fitsoln}) was obtained.
      Thus I am expecting a rather large error in this $T^c$ from the true transition temperature.
      Hence it is safer to reinterpret this $T^c$ as just a scale near the transition temperature.
      In units of this scale, there is no errors in temperature as the ratios of the temperatures are exact.
      The latter, $\beta_\ell^c=10$, is an extreme value and units of the horizontal axis, $T^h$,
      is a certain very high temperature.
      If one dares to extrapolate Equations~(\ref{eqn:fit},\ref{eqn:fitsoln}) to such an extreme,
      $T^h$ is about $10^{8}T^c$.
      At $\beta_\ell^c=10$, the spatial volume also is extremely small and one should worry about
      the finite spatial volume artifacts.
      The simulations are carried out for both $N_s=64$ and $96$ without showing statistically significant
      differences for the data of $4N_t<N_s$.
      See the main text for the fit curves.
  }}
  \label{fig:a0}
\end{figure}

The results are shown in Figure~\ref{fig:a0}.
Evidently, $\langle A_0\rangle\neq 0$ for the static gauge.
The fact is that $\langle A_0\rangle$ in the static gauge cannot be zero whatever the temperature is.
For given configurations (paths in the path integral), $\arccos(L)$ is a multivalued ill-defined ``function.''
To make sense out of this, the range must be restricted to $0\leq gA_0/T\leq\pi$.
Now, in order to achieve $\langle A_0\rangle = 0$, $A_0$ must fluctuate symmetrically around zero from configuration to configuration.
This, however, is impossible because the restriction always biases the fluctuations away from zero.%
\footnote{\label{ft:constraint} Similarly, $\langle L\rangle\neq 1$ can be argued.
  An $SU(2)$ element can be parameterized as $a_0\mathbf{1}_{2\times 2} + i \sum_{k=1}^3a_k\sigma_k$
  with the constraint $\sum_{k=0}^3a_k^2=1$.
  In this parameterization, $L=a_0$ and because of the constraint, this could never average to $1$.
  Also, since $a_0$ cannot all be $1$, $gA_0/T=\arccos(L)$ cannot all be $0$ either, hence
  $\langle A_0\rangle \neq 0$.
}
We can, therefore, categorically conclude that in the static gauge $\langle A_0\rangle\neq 0$ at any temperature.

Also visible in Figure~\ref{fig:a0} is the temperature dependence of $g\langle A_0\rangle/T$.
A quick examination shows that the decreases are slower than exponential but faster than logarithmic,
suggesting power-law falloffs.
Notice that in Equation~(\ref{eqn:pickScheme}), we could have chosen a slightly different scheme,
$C = c/\tilde{Z}(\tilde\beta_\ell)$ with a constant satisfying $0<c<1$.
This would have introduced an extra factor of $c^{T^c/T}$ in the renormalized $g\langle A_0\rangle/T - \pi/2$.
This factor would not introduce logarithmic behavior and $c^{T^c/T}\approx 1$ at high temperatures.
Thus the temperature dependence is different from the perturbative results
in which logarithmic falloff is expected through $g^2$.
So $\langle A_0\rangle \neq 0$ is a non-perturbative effect.

Let us examine the temperature dependence in more detail.
Because $SU(2)$ deconfinement transition is second order, $\langle A_0\rangle$ asymptotically approaches
$\pi/2$ as the temperature is lowered toward $T^c$, while Figure~\ref{fig:a0} suggests asymptotic falloff
toward zero as the temperature is increased.
Thus there is an inflection point at some temperature as can be seen in the left panel of Figure~\ref{fig:a0}
around $T/T^c=20$.
The right panel is in the range well beyond the inflection point.
I am interested in the falloff behavior beyond the inflection point.
For the data of the left panel with $T/T^c>20$, I have tried
\begin{equation}
  g\langle A_0 \rangle/T = \frac{a}{(\hat T + b)^n}
\end{equation}
where $\hat T:=T/T^c$, with the result $a=40(8)$, $b=41(4)$ and $n=0.815(37)$.
This is plotted in the left panel.
Despite its appearance, the fit is egregious
with the reduced $\chi^2$ about 150 (this is mainly due to the small errors of about $10^{-4}$)
and suggests that the true shape of the data is much more complicated.
Also this observation does not mean much as the fit parameters strongly depend on the scheme chosen.

By adjusting the scheme $\beta_\ell^c$, higher and higher temperature regions are examined.
I found that the power law slowly approaches to $n=1/2$ from above,
but doesn't quite get there; i.e., the falloff is slightly faster than $1/\sqrt{T}$.
This suggests that the falloff shape in an extremely high temperature region
is schematically $f(T)/\sqrt{T}$, where $f(T)$ is a slowly decreasing function.
At $\beta_\ell^c=10$, I have tried
\begin{equation}\label{eqn:fitlog}
  g\langle A_0 \rangle/T = \frac{a}{(\hat T + b)^{1/2}\big(1+c\ln(\hat T + b)\big)}
  \;,
\end{equation}
where $\hat T:=T/T^h$, with the result  $a=1.348(3)$, $b=0.407(6)$ and $c=0.044(1)$.
This is plotted in the right panel of Figure~\ref{fig:a0}.
The reduced $\chi^2$ is $17.5$; this is respectable considering the error bars of order $10^{-4}$ and
ignored systematic errors (see Appendix~\ref{app:details}).
I also observed that the parameter $b$ gets steadily smaller as higher ranges of temperature are explored.
There are other choices of the function $f(T)$ that yield comparable fit and it is difficult to
extract further physics out of the data.
I shun blind speculation; a more precise form of the data requires a good motivation from theoretical considerations.
\\

The observations made here are specific to $SU(2)$ and other gauge groups are not addressed.
When $N>2$, the $A_0$ expectation values certainly behave differently around the phase transition temperature,
for the order of the transition is different.
But given the behavior of the Polyakov loops for $N=3$,
it is reasonable to expect similar trend of the expectation values at high temperature.
It should, however, be explicitly verified that the falloff is {\it not} logarithmic,
i.e., not perturbative.

%%%%%%%%%%%%%%%%%%%%%%%%%%%%%%%%%%%%%%%%%%%%%%%%%%%%%%%%%%%%%%%%%%%%%
\subsection{Takeaways}\label{subsec:takeaways}

I have observed that the quantity $g\langle A_0\rangle/T$ for $SU(2)$ in the static gauge is non-vanishing
at all temperatures.
Here are the takeaways of this section:
\begin{itemize}
  \item for any allowed renormalization scheme,
    $g\langle A_0\rangle/T=\pi/2$ for $T\leq T^c$ and $0<g\langle A_0\rangle/T<\pi/2$ for all $T>T^c$;
  \item for any allowed renormalization scheme,
    $F(T):=\langle A_0\rangle/T$ in a high temperature range
    is a decreasing function of roughly $1/T$ to $1/\sqrt{T}$, in particular,
    it is not logarithmic implying the expectation value is a non-perturbative effect;
  \item the value of $F(T)$ at a temperature $T>T^c$ is scheme dependent and it is arbitrary
    in the range $0<gF(T)<\pi/2$.
\end{itemize}

In the temperature regime $T\ll T^c$, we still have $g\langle A_0\rangle/T=\pi/2$ all the way down to $T\to 0$.
The scale of the regime, however, is not $T$, but $\Lambda_\text{QCD}$.
Therefore, $\langle A_0\rangle/\Lambda_\text{QCD}$ and $T/\Lambda_\text{QCD}$ are both approximately zero
and the spacetime symmetry is restored.
In other words, $g\langle A_0\rangle/T\neq 0$ is not important in the temperature regime.
It becomes relevant at higher temperatures around $T^c\sim\Lambda_\text{QCD}$ and beyond.
When $T\gg T^c$, the function $F(T)$ may become very small in a scheme; however, this does not mean that
the expectation value is negligible because in another scheme, the value can be of order one.
Therefore, $F(T)$ generally cannot be ignored at high temperatures.

%%%%%%%%%%%%%%%%%%%%%%%%%%%%%%%%%%%%%%%%%%%%%%%%%%%%%%%%%%%%%%%%%%%%%
\section{Consequences and Discussion}\label{sec:cad}
%%%%%%%%%%%%%%%%%%%%%%%%%%%%%%%%%%%%%%%%%%%%%%%%%%%%%%%%%%%%%%%%%%%%%
I have shown that the groundstate of high temperature Yang-Mills theory
is nontrivial, dominated by the non-perturbative effects.
In this section, I am going to discuss consequences that follow and related issues.
\\
\\%
{\bf Perturbative and Non-Perturbative Contributions}

First I discuss contributions to a Yang-Mills thermal observable from different momentum ranges.
I am going to argue that zero temperature perturbative contribution comes from the momentum range $M\gg T$,
non-perturbative contribution from $M\lesssim T$ and possible thermal perturbative contribution
from the middle range.

In the momentum range $M\gg T$, the physics takes place at very short distances, much shorter than
the compactification size.
Then the process is insensitive to the temperature and the theory is in zero temperature perturbative regime,
i.e., $\langle A_0\rangle/M \approx 0$ and $T/M \approx 0$.
The contribution to a thermal observable from this range is the renormalization effect.
A good example is the first term on the right-hand side of Equation~(\ref{eqn:contoursum}).

In the range $M\lesssim T$, the dominant degrees of freedom are the zero modes; this is the range of EQCD$_3$.
Both $\langle A_0\rangle/M$ and $T/M$ are not negligible
and the non-perturbative structure of the thermal groundstate matters (provided that $T>T^c\sim\Lambda_\text{QCD}$).
It is natural to attempt perturbation around this groundstate;
this sort of computation was carried out by Nadkarni \cite{Nadkarni:1986as}.
The theory similar to the one shown in Equation~(\ref{eqn:L3D}) is perturbed with
unspecified expectation value of $A_0$.
At one-loop level, Nadkarni's tadpole consistency condition yields
a vanishing $A_0$ expectation value.
%\footnote{The tadpole condition is equivalent to determining the location of the minimum of
%the effective potential.}
This implies that the one-loop perturbation is inconsistent as elaborated in Section~\ref{sec:critique}.
At higher loops, the dimensionless quantity $g\langle A_0\rangle/T$ is computed in the series of $g^{2n}$ with
possibly vanishing coefficients.
Because there is no perturbative scale other than $T$ (and $gT$ for EQCD$_3$ but it must be remembered that $g$ is
not a small parameter: $g\sim 0.75$ at $T\sim 10^8T^c$),
the temperature dependence of $g\langle A_0\rangle/T$
comes only through $g$, i.e., logarithmic.
This is also inconsistent with the observation in Section~\ref{sec:lattice};
perturbation theory can only accommodate perturbatively modified groundstate from its trivial minimum.
The conclusion is that {\it Yang-Mills thermal perturbation theory is inconsistent with
the groundstate dominated by infrared non-perturbative effects.}
A compactified Yang-Mills theory is never under perturbative control.
Once we realize this, we do not really know how the coupling runs with temperature
and the whole assumption of small $\alpha_s$ at high temperature due to asymptotic freedom could collapse.
Thus we find that contributions to a thermal observable from the range $M\lesssim T$ are non-perturbative.

Finally let me discuss the mid-range momentum scale:
a gray zone between the perturbative and non-perturbative ranges.
The relevant degrees of freedom are the heavy modes and hard zero modes.
Write $F(T)=\langle A_0\rangle/T$ as in Section~\ref{subsec:takeaways}.
Consider a heavy positive $n$ mode.
We then have the approximate scale $M=nT$ or $T/M=1/n$, leading to $\langle A_0\rangle/M=F(T)/n$.
When $n$ is very large, say $10^6$, this mode should yield a near perturbative contribution as discussed
for the range $M\gg T$, while $n=1$ a non-perturbative one.
Thus the heavy mode contributions consist of both perturbative and non-perturbative effects.
Put it differently, thermal heavy mode perturbation, like Equation~(\ref{eqn:cosineexpansion}),
can account for part of the heavy mode contributions but certainly not all.
As for the hard zero modes, it is difficult to unravel the contributions from different momentum scales
as discussed in Section~\ref{sec:effS}.
Hence it is more natural to interpret the contributions from the hard zero modes as renormalization effects
on EQCD$_3$ so that they are part of non-perturbative 3D contributions.

Overall, a mix of perturbative and non-perturbative contributions comprises a Yang-Mills thermodynamic quantity.
The sizes of the ranges discussed above do not translate to the size of the contribution.
It may appear that the perturbative momentum range is rather restricted but this does not mean its
contribution is necessarily small, while
neither does high temperature guarantee perturbative dominance.
Thus it is important to disentangle these different contributions in lattice thermodynamic observables.
It may be possible to find observables that are mostly perturbative or non-perturbative.
Or there may be an observable that somehow allows the lattice professionals to clearly separate
the contributions coming from different momentum regions.
\\
\\
{\bf 3D Effective Theory}

The discussion above makes it clear that QGP as a nearly free gas of gluons (and quarks) is very unlikely.
What is more likely is that the elementary particle picture itself is not available -- especially for a 3D description --
because of the non-perturbative nature of the thermal equilibrium.
It is, therefore, very difficult to come up with a useful form of 3D effective theory.
Unlike $\chi$PT, there is no useful symmetry in sight and crucially, a dearth of pertinent experimental data
hinders the identification of useful 3D degrees of freedom.
Even if we decide to try EQCD$_3$ of the form Equation~(\ref{eqn:L3D}), we would have to match
the coefficients non-perturbatively to data from experiments or lattice.
At the moment, this is not a viable path to take and moreover, EQCD$_3$ would not be a simple theory
anyway requiring full non-perturbative treatments.
If a 3D effective theory of high temperature Yang-Mills theory -- as useful as $\chi$PT --
exists then it must be described by interesting degrees of freedom radically different from gluons.
It would be exciting if the professionals are able to come up with such a theory.
\\
\\
{\bf Confinement}

Let me now turn to the most interesting physics in the context.
I have often stated that the 3D theory ``confines'' but this has nothing to do
with the 4D confinement.
In the literature, this is most often referred to as the confining property of MQCD$_3$ --
$SU(N)$ Yang-Mills theory in three dimensions.
There is no doubt that MQCD$_3$ confines but preceding sections strongly suggest that 
it is an unlikely low energy candidate.
There is, however, fascinating evidence that a spatial 3D effective theory should ``confine'':
the area law behavior of a large {\it spacelike} Wilson loop, below {\it and above} $T^c$.
The associated coefficient of the exponential decay is misleadingly dubbed ``spatial string tension.''
This has nothing to do with the string tension and confinement, but is a particular feature of
the vacuum fluctuations measured by the Wilson loop.

It seems that Borgs in Reference~\cite{Borgs:1985qh} is the first to methodically consider this topic
and has analytically shown the spatial area law on the high temperature lattice.
The first definitive lattice simulations of the phenomenon were carried out by Bali {\it et al.}~\cite{Bali:1993tz}.
They observed that the ``spatial string tension'' coincides with the string tension well below $T^c$,
then as the temperature nears $T^c$, the latter drops sharply towards zero while the former shows little change.
As the temperature rises beyond $T^c$, the ``spatial string tension'' goes up and this increase is not incompatible
with a naive expectation that the square root of the ``spatial string tension'' is proportional to $g_3^2=g^2T$.
What is striking here is the uneventful behavior of the spatial area law across the $\mathbb{Z}_N$ phase transition.
It is as if the ``confining'' mechanism is insensitive to the $\mathbb{Z}_N$, which plays a prominent role
in the study of confinement.

\begin{figure}[h]
  \centering
  \includegraphics{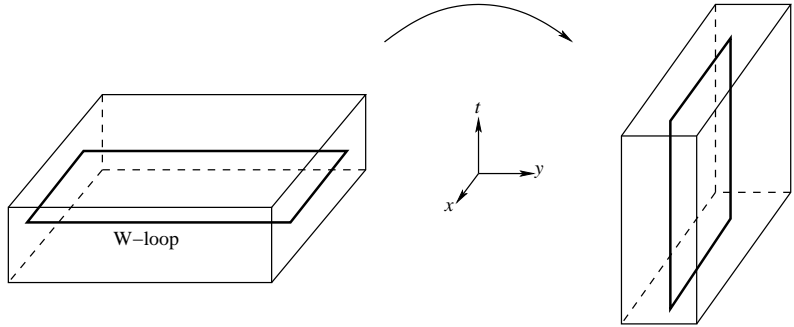}
  \caption{\footnotesize{Reinterpretation of the coordinates.
  }}
  \label{fig:rorate}
\end{figure}

Before going further, I remedy the problem that the reader and I are both tired of
the quotation marks around the words already.
This only requires simply reinterpreting the compactified time direction as a spatial coordinate.
This is a zero temperature system with one space compactified.
And the spacelike Wilson loop discussed so far is recast to a timelike loop whose spatial extent is
in one of the uncompactified directions. See Figure~\ref{fig:rorate}.
In this way ``confinement'' becomes bona fide confinement.
The discussion so far implies that the new system is {\it always} confined whether or not
the $\mathbb{Z}_N$ symmetry -- associated with the compactified spatial direction -- is broken,
and shows little response at the critical point.

How do we explain this?

If a tightly compactified Yang-Mills vacuum were perturbative, then perturbation theory would have to
be able to describe confinement; this is a false assumption.
In this work, I have found that Yang-Mills theory is non-perturbative,
even when the compactified dimension is smaller than the critical size.
So the persistent confinement would appear to be consistent,
but my observation can hardly say anything about confinement.
One could try to argue as follows:
1) Suppose that observed $\langle A_0 \rangle \neq 0$ breaks the $SU(2)$ gauge symmetry to $U(1)$ and
the theory is effectively three-dimensional Georgi-Glashow model;
2) Suppose that the massive $A_0$ scalar and $W^\pm$ (the off-diagonal matters) can be ``integrated out'';
3) Further suppose that the resulting instantons (the monopoles) are dilute;
4) Then the ``conjecture'' is that my system confines \'a la Polyakov~\cite{Polyakov:1976fu}.
This is nothing but wishful thinking.
For the point 1), the spontaneous symmetry breaking is plausible but the 3D Georgi-Glashow model as
an effective theory is hard to sell.
There are many more possible terms and none of the coefficients in the low energy theory can be determined
because the full and effective theories are not perturbative.
For the point 2), it is very often said that the $W$s can be integrated out leaving no trace, but this cannot be true.
Especially in a non-abelian gauge theory, heavy fields do not simply decouple but generate infinitely many
operators in the effective action.
They will, in particular, alter the shape of the very scalar potential that is giving
the expectation value and masses to the fields, potentially violating the self-consistency.
Also as pointed out in Reference~\cite{Ambjorn:1998qp}, $W$s can screen test charges.
Then after exactly integrating out the $W$s, the theory must somehow manage to demonstrate the screening of the test charges
by the $W$s that do not exist in the effective theory.
This means that the integration generates highly non-trivial relevant operators in the effective action.
In short, heavy fields, especially $W$s, may not just be integrated out into thin air.
The point 3) is not possible to justify at all because, as mentioned above,
the coefficients of the effective theory is unknown.
The only sure thing is the existence of the topological charges $\pi_2(SU(2)/U(1))$,
but this does not guarantee the existence of a soliton solution.

Just for the sake of argument, suppose that the 3D monopole instantons are confining the system
when the compactification is extremely small.
As the compactified dimension is gradually opened up, the instantons start to draw worldlines in spacetime.
When the compactification is still small, it is energetically favorable for the worldlines to wrap around
the compact coordinate.
But at some larger size, the monopoles and antimonopoles make small topologically trivial loops of worldlines.
When the loops intersect with the the minimal surface bounded by the Wilson loop, the intersections are
in tight pairs of monopoles.
This destroys the magnetic disorder and the Wilson loop would develop perimeter law; the small loops can
link to the Wilson loop only at the edges.
This is a contradiction since we discussed that the system is always confined.
So instead of this happening at the critical size of the compact dimension, an entirely different phenomenon --
like the monopole condensation -- must smoothly ``usurp'' the confinement mechanism, presumably thanks to strong coupling.
Let us now consider in opposite way by gradually compactifying a spatial dimension.
When the size is very large, Yang-Mills theory confines and
let us suppose that this is a universe of dual superconductor
in which the monopole condensate permeates the spacetime.
As the size of the compact dimension is reduced and crosses the critical point, confinement persists;
then it would be natural to assume that the condensate persists as well.
This, however, cannot be the case because we know that if we reinterpret the compact direction as time,
the condensate would dissolve at the critical size (the system deconfines);
this is the same for the spatial compactification in the Euclidean setup.
Then at the critical size and beyond, another mechanism, like the monopole instantons described above,
must confine the system.
So as long as one insists on the monopole condensate/instanton confinement mechanisms,
the very awkward confinement usurpation must take place smoothly.
This can describe the confinement across the $\mathbb{Z}_N$ transition, but not a very attractive one
unless there is a spectacular relationship between the monopole instantons and the monopole condensate
that I am unaware of.

There is a better alternative: the center vortex confinement mechanism
(see Greensite's book \cite{Greensite:2011zz} and Reference~\cite{Engelhardt:1999fd}).
In this picture, the Yang-Mills vacuum is two-dimensional, closed center vortex
surfaces superimposed on a non-confining configuration.
Confinement occurs when the vortex surfaces ``percolate'' spacetime, so that
the vortices randomly pierce the Wilson loop surface in a topologically nontrivial way,
causing the magnetic disorder.
The (thin) vortex configurations are observed in the lattice at finite temperatures by Engelhardt {\it et al.}
in Reference~\cite{Engelhardt:1999fd}
and we can freely reinterpret the system as our spatially compactified zero temperature setup.
The observations are made in three dimensional slices so that the vortices appear as one dimensional loops.
When the $\mathbb{Z}_N$ symmetry is intact, the vortex lines percolate spacetime isotropically.
But when the symmetry is broken, the vortices cease to percolate in a three-dimensional slice
that include the compactified direction.
The vortex lines tend to align with and wrap around the compactified direction, only fluctuating little in
the transverse directions.
Meanwhile, in a slice that does not include the compactified dimension, vortices continue to percolate
into the $\mathbb{Z}_N$ broken phase without noticeable difference across the phase transition.
In other words, the vortex configuration in the uncompactified three dimensions  does not change qualitatively
and, in our system, the confinement mechanism is identical across the $\mathbb{Z}_N$ transition
associated with the compactified spatial direction.
This sort of anisotropy is hard to imagine for a Euclidean universe of dual superconductor.
(The spacetime permeating monopole condensate would have to dissolve when the time coordinate is compactified
down to the critical size and the condensate would somehow have to persist
when a spatial coordinate is compactified at any size.)
The vortex picture is attractive but difficult to relate to my work.
My observation is that the compactified component of the gauge field acquires nonzero expectation value
at any size.
This breaks the $SU(2)$ to $U(1)$, so this can partially explain
the abelian dominance \cite{Suzuki:1989gp,Ejiri:1994uw},
but not the center dominance \cite{DelDebbio:1996lih,Greensite:2011zz} in the string tension.
The static gauge is not adequate in filtering vortices out of the vacuum.

The center vortex confinement mechanism is a promising scenario but the dynamics and shape
of the flux tube are hard to visualize.
In our setup, the periodically identified compact direction is one of the transverse directions
of a flux tube.
So the flux tube is self-interacting with infinite series of its own images in its entire length.
The lattice measurements of string tension suggest rather complicated interaction.
When the size of the compact dimension, $L$,  is larger than the critical size
($L>L_c\sim 1/\Lambda_\text{QCD}$),
the linear energy density of the flux tube is relatively insensitive to $L$.
At $L\sim L_c$, the flux tubes are touching each other or nearly so as the thickness of the tube is presumably
about $L_c\sim 1/\Lambda_\text{QCD}$.
As the size is further reduced, the linear energy density increases roughly as $1/L^2$.
This interesting behavior of the flux tube has not been studied except its tensions.
The chromoelectric field can be measured on the lattice by probing a plaquette around a large Wilson loop,
and it is very interesting to see how the shape and profile of the
``squeezed'' flux tube would respond to the size $L$.
I have run several simulations but the results are inconclusive; the profile significantly depends on
the type of smearing used and the continuum limit is not addressed.
This is left for professionals with resources and will.
\\
\\%
{\bf Chiral Symmetry Breaking}

Let me continue on the spatially compactified setup.
We have discussed that the system is non-perturbative and confining at small sizes of compact dimension
beyond the $\mathbb{Z}_N$ critical point.
Presence of (fundamental) fermions makes it difficult to define confinement \cite{Greensite:2011zz}
but the behavior of fermions in otherwise confining vacuum can be studied.
To this end, it would be very interesting to see whether $\chi$SB also persists
across the spatial $\mathbb{Z}_N$ transition.
(The $\mathbb{Z}_N$ transition would no longer be non-analytic but the spatial Polyakov loop should still
show signs of a smooth crossover transition, such as a peak in the susceptibility.)
Because $\chi$SB may not be directly related to gluodynamics (think of NJL),
the chiral condensate may have different behavior across the phase transition.
Thus, the spatially compactified system here is not just an academic curiosity
but it could shed some lights on the relation between confinement and $\chi$SB, or lack of it.

At finite temperature, the fermions are necessarily anti-periodic in the time direction.
This is no longer the case for a spatially compactified system: fermions can have ``twists''
at the boundary, including a $U(1)$ phase.

When the fermions are endowed with anti-periodic boundary conditions, the system is the same as
the finite temperature counterpart.
Then, in this case, the chiral restoration must occur around the (crossover) $\mathbb{Z}_N$
transition point.
In this chiral-symmetric phase, the system still confines in the sense that the flux tubes
form up to the breaking point.
This is a novel situation where confinement and $\chi$SB are completely disentangled.
This may suggest that they are separate phenomena caused by separate mechanisms.

When the fermions are periodic in the compactified spatial direction, one may not reinterpret
the system as finite temperature setup.
For this case, I do not know how chiral condensate behaves across the transition: $\chi$SB
may persist or chiral symmetry may be restored.
A boundary condition is a non-local restriction but it changes the momentum spectrum.
For the periodic boundary condition, there are fermionic zero modes at any compactification size.
Then if $\chi$SB turns out to persist, it is an indication of infrared dynamics
that is responsible for $\chi$SB.

These are speculations but totally falsifiable.
Further study must be carried out, especially on the lattice.
I hope that a good professional or two will take up the subject and expand our knowledge of nature.
\\

%%%%%%%%%%%%%%%%%%%%%%%%%%%%%%%%%%%%%%%%%%%%%%%%%%%%%%%%%%%%%%%%%%%%%%%%
\section*{Acknowledgments}
%%%%%%%%%%%%%%%%%%%%%%%%%%%%%%%%%%%%%%%%%%%%%%%%%%%%%%%%%%%%%%%%%%%%%%%%
First and foremost, I would like to thank the operators of {\tt arXiv} and {\tt sci-hub}, without
whom this work could not have been completed.
Kudos to those who let their papers openly accessible to the public, like Utrecht University Library.
I also thank for the hospitality of Ehime Prefectural Library where Mx. MiCan encouraged me to wrap up
this 12-year-old project already and obeying them Section~\ref{sec:cad} was written.

\bigskip
\bigskip

%%%%%%%%%%%%%%%%%%%%%%%%%%%%%%%%%%%%%%%%%%%%%%%%%%%%%%%%%%%%%%%%%%%%%
%%%%%%%%%%%%%%%%%%%%%%%%%%%%%%%%%%%%%%%%%%%%%%%%%%%%%%%%%%%%%%%%%%%%%
\appendix
%%%%%%%%%%%%%%%%%%%%%%%%%%%%%%%%%%%%%%%%%%%%%%%%%%%%%%%%%%%%%%%%%%%%%
%%%%%%%%%%%%%%%%%%%%%%%%%%%%%%%%%%%%%%%%%%%%%%%%%%%%%%%%%%%%%%%%%%%%%

%%%%%%%%%%%%%%%%%%%%%%%%%%%%%%%%%%%%%%%%%%%%%%%%%%%%%%%%%%%%%%%%%%%%%
\section{BRST Symmetry under Static Gauges}\label{app:BRST}
%%%%%%%%%%%%%%%%%%%%%%%%%%%%%%%%%%%%%%%%%%%%%%%%%%%%%%%%%%%%%%%%%%%%%
Consider a gauge fixed Lagrangian
\begin{equation}\label{eqn:gfixedL}
  \mathcal{L} = \frac{1}{4}\big( F_{\mu\nu}^a \big)^2 - \frac{\xi}{2}\big( B^a \big)^2  + B^a V_\mu A_\mu^a
  - \bar c^a V_\mu \big( \partial_\mu c^a - gf^{abc} A_\mu^b c^c \big)
  \;.
\end{equation}
The auxiliary filed $B^a$ can be eliminated to yield a gauge fixing term $(1/2\xi)(V_\mu A_\mu^a)^2$.
The multiplet $V_\mu$ can be chosen for specific gauges; for example, $V_\mu=\partial_\mu$ gives the general Lorentz gauge
and $V_\mu=(\lambda,\vec 0)$ with $\lambda\to\infty$ gives an axial (the temporal) gauge $A_0=0$.
The last term in the Lagrangian is the corresponding ghosts.
This is invariant under the BRST transformation
\begin{eqnarray}
  \delta A_\mu^a(t,\vec x) &=& \epsilon\partial_\mu c^a(t,\vec x) - \epsilon g f^{abc}A_\mu^b(t,\vec x)c^c(t,\vec x)
  \nonumber\\
  \delta c^a(t,\vec x) &=& -\frac{1}{2}\epsilon g f^{abc} c^b(t,\vec x)c^c(t,\vec x)
  \nonumber\\
  \delta\bar c^a(t,\vec x) &=& \epsilon B^a(t,\vec x)
  \nonumber\\
  \delta B^a(t,\vec x) &=& 0
  \;,
\end{eqnarray}
where $\epsilon$ is an anti-commuting global transformation parameter.
Notice that the gauge field part is exactly a gauge transformation in which $\epsilon c^a(t,\vec x)$ is
playing the role of the transformation parameter.
Therefore, the split (\ref{eqn:split}) is not respected by the BRST symmetry either -- in general, that is.

The static gauge (\ref{eqn:staticgauge})
is a very special and perhaps unique gauge that evades the problem.
To see this, consider a static gauge defined by
\begin{equation}
  V_\mu = (\lambda\partial_0,V_i)
  \quad\text{with}\quad
  \lambda\to\infty \;,
\end{equation}
where the spatial gauge fixing $V_i$ need not be specified for this general discussion here.
By examining the terms in the Lagrangian (\ref{eqn:gfixedL}) with this gauge, one finds that
$\partial_0A_0^a=0$ as well as $\partial_0c^a=0$, $\partial_0\bar c^a=0$ and $\partial_0B^a=0$
(some require integration by parts and this is well-defined thanks to the periodic boundary conditions),
leaving the Lagrangian in the form
\begin{equation}\label{eqn:statGLag}
  \mathcal{L} = \frac{1}{4}\big\{ F_{\mu\nu}^a(t,\vec x) \big\}^2
  - \frac{\xi}{2}\big\{ B^a(\vec x) \big\}^2  + B^a(\vec x) V_i \hat A_i^a(\vec x)
  - \bar c^a(\vec x) V_i \big\{ \partial_i c^a(\vec x) - gf^{abc} \hat A_i^b(\vec x) c^c(\vec x) \big\}
  \;.
\end{equation}
The first term involves the full 4D field $A_\mu^a(t,\vec x)$ but in the rest,
I have dropped the products of orthogonal Fourier modes; namely, I wrote $\hat A_i^a(\vec x)$
instead of $\hat A_\mu^a(\vec x) + A_\mu'^a(t,\vec x)$.
Notice that the heavy mode $A_\mu'^a(t,\vec x)$ has decoupled from the ghosts, as mentioned
in Section~\ref{sec:effS}.

The Lagrangian (\ref{eqn:statGLag}) is invariant under the BRST transformation
\begin{eqnarray}
  \delta \big\{\hat A_\mu^a(\vec x) + A_\mu'^a(t,\vec x)\big\} &=& \epsilon\partial_\mu c^a(\vec x)
  - \epsilon g f^{abc} \big\{\hat A_\mu^b(\vec x) + A_\mu'^b(t,\vec x)\big\} c^c(\vec x)
  \nonumber\\
  \delta c^a(\vec x) &=& -\frac{1}{2}\epsilon g f^{abc} c^b(\vec x)c^c(\vec x)
  \nonumber\\
  \delta\bar c^a(\vec x) &=& \epsilon B^a(\vec x)
  \nonumber\\
  \delta B^a(\vec x) &=& 0
  \;.
\end{eqnarray}
I can, therefore, define a formal BRST transformation that respects the split (\ref{eqn:split}):
\begin{eqnarray}\label{eqn:formalBRST}
  \delta\hat A_0^a(\vec x) &=&  - \epsilon g f^{abc} \hat A_0^b(\vec x) c^c(\vec x)
  \nonumber\\
  \delta \hat A_i^a(\vec x) &=& \epsilon\partial_i c^a(\vec x)
  - \epsilon g f^{abc} \hat A_i^b(\vec x) c^c(\vec x)
  \nonumber\\
  \delta A_i'^a(t,\vec x) &=&  - \epsilon g f^{abc} A_i'^b(t,\vec x) c^c(\vec x)
  \nonumber\\
  \delta c^a(\vec x) &=& -\frac{1}{2}\epsilon g f^{abc} c^b(\vec x)c^c(\vec x)
  \nonumber\\
  \delta\bar c^a(\vec x) &=& \epsilon B^a(\vec x)
  \nonumber\\
  \delta B^a(\vec x) &=& 0
  \;.
\end{eqnarray}
Notice that the gauge field parts again are gauge transformations but now corresponding to
a {\it three dimensional spatial} gauge symmetry,
under which $\hat A_0$ and $A_i'$ are transforming as adjoint matters.

Thus far, we have seen that the static gauge allows for the split (\ref{eqn:split}) and a well-defined BRST symmetry.
Under the reasonable assumption that the integration measure of $A_i'$ is invariant under the transformation (\ref{eqn:formalBRST}),
the result of the integration over the heavy modes, call it $\mathcal{L}^\text{eff}$, will be BRST invariant.
The part in $\mathcal{L}^\text{eff}$ comprising purely $\hat A_\mu$ is invariant under the 3D gauge symmetry just noted.
The gauge fixing term, $(1/2\xi)(V_i\hat A_i)^2$, is not affected by the integration over $A_i'$, so the terms that involve
the $B$ field is unaffected.
The ghosts do not directly couple to $A_i'$ so the integration over the field cannot affect their form in the Lagrangian either.
Similar logic can be applied to show there is no new vertices in $\mathcal{L}^\text{eff}$ that couple $c$ and $\hat A_i$.
After all, $\mathcal{L}^\text{eff}$ takes the form of $\mathcal{L}_\text{3D}$ in Equation~(\ref{eqn:L3D}),
but now the ``dots'' include the specific gauge fixing and the corresponding ghost terms
shown in Equation~(\ref{eqn:statGLag}).
I emphasize again that this conclusion is specific to the static gauge.

%%%%%%%%%%%%%%%%%%%%%%%%%%%%%%%%%%%%%%%%%%%%%%%%%%%%%%%%%%%%%%%%%%%%%
\section{Derivation of $V_\text{3D}$}\label{app:bosonDet}
%%%%%%%%%%%%%%%%%%%%%%%%%%%%%%%%%%%%%%%%%%%%%%%%%%%%%%%%%%%%%%%%%%%%%
The task here is to evaluate the integral
\begin{eqnarray}
  V_\text{3D} & = & \frac{1}{L_s^3}{\sum_{m,n}}'\tr\ln\big[ (p_0-g\lambda_{mn})^2+(p_i)^2 \big]
  \nonumber\\
  & = & \frac{1}{L_s^3}{\sum_{m,n}}'\int\frac{d^3p}{(2\pi)^3} I
  \quad\text{with}\quad
  I := T{\sum_l}' \ln\big[ (p_0-g\lambda_{mn})^2+p^2 \big]
  \;,
\end{eqnarray}
where I have defined $p^2:=p_ip_i$, $p_0=2\pi lT$ and the prime on the $l$-sum indicates that the zero mode is not included.
Recall the familiar formula
\begin{equation}\label{eqn:contoursum}
  T\sum_{l=-\infty}^{\infty}f(2\pi lT) = \int\frac{dp_0}{2\pi}\frac{f(p_0)+f(-p_0)}{2}
  + \int_{-\infty + i\epsilon}^{+\infty + i\epsilon}\frac{dp_0}{2\pi}\frac{f(p_0)+f(-p_0)}{e^{-ip_0/T}-1}
  \;,
\end{equation}
where $f(p_o)$ is an arbitrary function that is analytic in the neighborhood of the real axis,
the second term of the right-hand side is a contour integral and the large semi-circle must
be closed above.
Using this formula, one obtains
\begin{eqnarray}\label{eqn:dIdp2}
  \frac{dI}{dp^2} &=& T{\sum}'\frac{1}{(p_0-g\lambda_{mn})^2+p^2}
  \nonumber\\
  &=& \int\frac{dp_0}{2\pi}\frac{1}{p_0^2+p^2}
  +T\frac{d}{dp^2}\bigg\{ \ln\big( 1- e^{(ig\lambda_{mn}-\sqrt{p^2})/T} \big)
  + \ln\big( 1- e^{(-ig\lambda_{mn}-\sqrt{p^2})/T} \big) \bigg\}
  \nonumber\\
  &-& \frac{T}{p^2+(g\lambda_{mn})^2}
  \;,
\end{eqnarray}
where the first term in the second line is the $\lambda_{mn}$-independent zero temperature contribution
and the last term is the zero mode subtraction.
The integration over $p^2$ in Equation~(\ref{eqn:dIdp2}) can be carried out trivially and I am going to drop the $\lambda_{mn}$- and
$p^2$-independent terms:
\begin{eqnarray}
  \int\frac{d^dp}{(2\pi)^d}I &=& -T\int\frac{d^dp}{(2\pi)^d}\sum_{k=1}^\infty \frac{1}{k}
  \big( e^{k(ig\lambda_{mn}-\sqrt{p^2})/T} + e^{k(-ig\lambda_{mn}-\sqrt{p^2})/T} \big)
  \nonumber\\
  && - T\int d(g\lambda_{mn})^2\int\frac{d^dp}{(2\pi)^d}\frac{1}{p^2+(g\lambda_{mn})^2}
  \nonumber\\
  &=& -2T\sum_k\frac{\cos(kg\lambda_{mn}/T)}{k}\int\frac{d^dp}{(2\pi)^d}e^{-k\sqrt{p^2}/T}
  + \frac{T}{4\pi}\int d(g\lambda_{mn})^2\sqrt{(g\lambda_{mn})^2}
  \nonumber\\
  &=& -\frac{2T^4}{\pi^2}\sum_{k=1}^{\infty}\frac{\cos(kg\lambda_{mn}/T)}{k^4}
  + \frac{T}{6\pi}|g\lambda_{mn}|^3
  \;,
\end{eqnarray}
where $d\to3$ and the linear divergence of the zero mode subtraction term was dimensionally regularized.
From the Table of Integrals \cite{IntegralTable} p.47, we have a remarkable relation
\begin{equation}\label{eqn:thesum}
  \sum_{k=1}^{\infty} \frac{\cos(kx)}{k^4} = \frac{\pi^4}{90} - \frac{\pi^2 [ x ]^2}{12}
  + \frac{\pi [x]^3}{12} - \frac{[x]^4}{48}
  \quad\text{where}\quad
  [x] := x \bmod 2\pi
  \;.
\end{equation}
Then after trivially transforming the expression in the four dimensional momentum space back to
the four dimensional position space, i.e., supplying the spacetime volume, I get
\begin{eqnarray}
  V_\text{3D} = \frac{2T^3}{\pi^2}{\sum_{m,n}}'\bigg[\bigg\{ -\frac{\pi^4}{90} + \frac{\pi^2 [g\lambda_{mn}/T]^2}{12}
  - \frac{\pi [g\lambda_{mn}/T]^3}{12} + \frac{[g\lambda_{mn}/T]^4}{48} \bigg\}
  +  \frac{\pi |g\lambda_{mn}/T|^3}{12} \bigg]
  \;.
\end{eqnarray}

%%%%%%%%%%%%%%%%%%%%%%%%%%%%%%%%%%%%%%%%%%%%%%%%%%%%%%%%%%%%%%%%%%%%%
\section{Simulation Details}\label{app:details}
%%%%%%%%%%%%%%%%%%%%%%%%%%%%%%%%%%%%%%%%%%%%%%%%%%%%%%%%%%%%%%%%%%%%%
Versions of a CUDA C++ code were run on a consumer grade video accelerator (GeForce RTX 2060).
Parallel algorithms were implemented as much as possible to take advantage of the parallel computing device;
in particular, the lattices were thermalized in the checkerboard manner.
Pseudorandom numbers were generated on the device using the cuRAND library functions:
{\tt curand\_init()} and {\tt curand\_uniform\_double()}.

The standard Wilson action is employed for $SU(2)$ pure gauge theory.
I have used Creutz's heatbath algorithm \cite{Creutz:1980zw} with the modification of Kennedy and Pendleton \cite{Kennedy:1985nu}.
One heatbath sweep is followed by 4 overrelaxation steps \cite{Creutz:1987xi}.
In the following, I will call this 5-step thermalization sweep ``a compound sweep.''

The lattice size is measured in units of lattice spacing where $N_t$ and $N_s$ are temporal and spatial extents, respectively.
Because of the checkerboard algorithm, $N_{t,s}$ are even integers.
The lattice is in the form $N_t\times N_s^3$ and all four directions are periodically identified.
The temporal extent $N_t$ is varied from 4 to as large as 22 and $N_s$ is fixed at either 64 or 96 such that
$4N_t<N_s$ is always satisfied.
The size $4N_t=N_s$ is avoided because I found this pretty near the border of finite spatial volume independence.
I eschew $N_t=2$ because of discretization pollution; see below for more explanation.

Before measurements are taken, 200 compound sweeps are carried out.
Each measurement is separated by 5 compound sweeps.
I found this enough to reduce the autocorrelation of the Polyakov loops, except in the vicinity of the phase transition point.
To avoid the possible contamination by the critical slowing down,
the data nearest to the transition point are not used in the analyses.

The statistical errors are estimated in the standard fashion and no systematic errors are taken into account.
Several runs are carried out for both $N_s=64$ and $96$ in otherwise identical settings.
I have confirmed that there are no statistically significant differences for the data points satisfying $4N_t<N_s$.

Thanks to the space averaging of the Polyakov loop or the arc-cosine of it,
the statistical errors are relatively small, and thus fewer configurations, 400, are collected for each expectation value.

Nonlinear curve fitting is done through {\tt gnuplot}'s {\tt fit} command, which employs the Marquardt-Levenberg method
for the (weighted) least squares.
The errors in the fit parameters reported in this command are $\hat\chi^2$-improved version of them, in which
the original variances of the data are multiplied by the reduced $\chi^2$ of the fit curve, then in turn,
this improved variances are used in the computation of the error matrix.%
\\

For the rest of this appendix, I discuss the discretization pollution mentioned above.
In (continuum) thermal field theory, renormalization is done at zero temperature and no further renormalization
is required at nonzero temperature.
This is because the renormalization is an ultraviolet treatment and insensitive to the compactified time direction.
In other words, the temperature should be irrelevant to ultraviolet processes.
On the lattice, then, Wilson loops of the size comparable to the lattice spacing -- such as the elementary
plaquette -- should be insensitive to changing temperature.
If the temperature comes too close to the lattice cutoff scale ($N_t$ too small)
the plaquette expectation values may develop anisotropy and show dependence on the temperature.
One must expect that the thermal observables at such small $N_t$ are polluted by the lattice
cutoff and discretization artifact.

This must be tested at a fixed $\beta_\ell$.
Suppose that $a(\beta_1)=2a(\beta_2)$.
Then a lattice with $N_t$ at $\beta_\ell=\beta_1$ and one with $2N_t$ at $\beta_\ell=\beta_2$
are at the same temperature.
But the plaquette expectation values are different on these lattices because the observable does not scale.
To circumvent this issue, the test must be done by varying $N_t$ at a fixed $\beta_\ell$, rather than
varying $\beta_\ell$ at a fixed $N_t$.

\begin{figure}[t]
  \centering
  \includegraphics{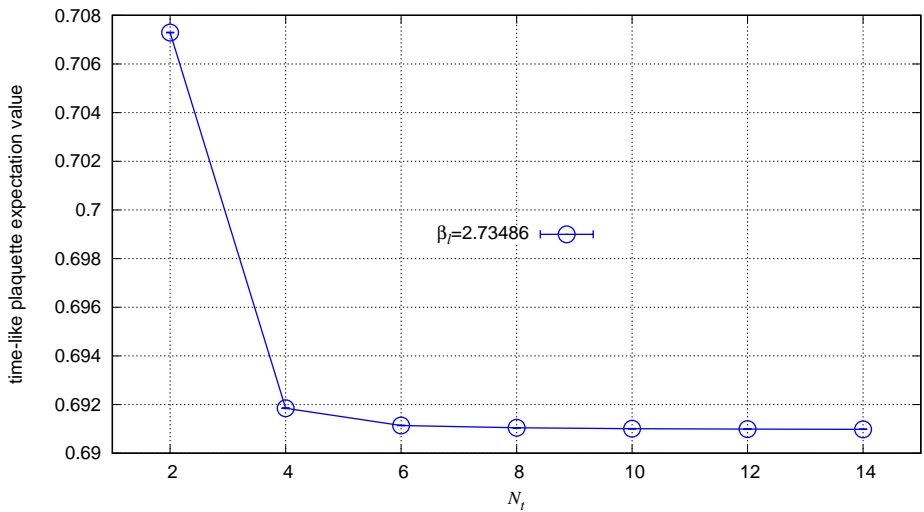}
  \caption{\footnotesize{The expectation values of timelike plaquette against $N_t$,
      the size of the lattice in time direction.
      The spatial size is fixed at $N_s=64$ and the coupling is set at $\beta_\ell=2.73486$.
      This coupling corresponds to $N_t^c=16$ so the system is in deconfined phase,
      but the plaquette expectation value is insensitive to the phase transition as long as
      the coupling is fixed.
    }}
  \label{fig:plq_t16}
\end{figure}

Figure~\ref{fig:plq_t16} shows the expectation values of timelike plaquette against $N_t$.
The expectation values stay insensitive to the temperature down to about $N_t=6$ or $8$,
but at $N_t=2$ the expectation value is clearly affected by the temperature.
It is not shown but the spatial plaquette expectation values behave very differently; below $N_t=6$ or $8$,
they split from the timelike counterparts and goes downward in the figure, though less steeply so, leaving
the total spacetime plaquette expectation values go up.
This observation clearly exposes the anisotropy.
This strongly suggests that the Polyakov loops measured at $N_t=2$ are contaminated by the discretization artifact.
I have decided to use $N_t=4$ and $6$ in need of data points but they must be treated with caution.

%\pagebreak

%%%%%%%%%%%%%%%%%%%%%%%%%%%%%%%%%%%%%%%%%%%%%%%%%%%%%%%%%%%%%%%%%%%%%

\end{document}